\documentclass[twocolumn,superscriptaddress]{revtex4}
\usepackage{epsfig}
\usepackage{amssymb,amsfonts,amsmath}
\usepackage[pdftex]{pstricks}
\include{epsf}

\newcommand{\ben}{\begin{enumerate}}
\newcommand{\een}{\end{enumerate}}

\newcommand{\<}{\langle}
\renewcommand{\>}{\rangle}

\newcommand{\beq}{\begin{equation}}
\newcommand{\eeq}{\end{equation}}
\newcommand{\bea}{\begin{eqnarray}}
\newcommand{\eea}{\end{eqnarray}}
\newcommand{\fig}[1]{Fig.~\ref{#1}}
\usepackage{amsmath}
\usepackage{amssymb}
\usepackage{amsthm}
\usepackage{amsfonts}

\begin{document}
\title{Dynamical criticality in the collective activity of a population of retinal neurons}
\author{Thierry Mora}
\affiliation{Laboratoire de physique statistique, \'Ecole normale sup\'erieure, CNRS and UPMC, 24 rue Lhomond, 75005 Paris, France}
\author{St\'ephane Deny}
\affiliation{Institut de la Vision, INSERM and UMPC, 17 rue Moreau, 75012 Paris, France}
\author{Olivier Marre}
\affiliation{Institut de la Vision, INSERM and UMPC, 17 rue Moreau, 75012 Paris, France}

\date{\today}
\linespread{1}

\begin{abstract}
Recent experimental results based on multi-electrode and imaging techniques have reinvigorated the idea that large neural networks operate near a critical point, between order and disorder \cite{Chialvo:2010p13201,Mora:2011p12729}. 
However, evidence for criticality has relied on 
the definition of arbitrary order parameters, or on models that do not address the dynamical nature of network activity.
Here we introduce a novel approach to assess criticality that overcomes these limitations, while encompassing and generalizing previous criteria. 
We find a simple model to describe the global activity of large populations of ganglion cells in the rat retina, and show that their statistics are poised near a critical point. 
Taking into account the temporal dynamics of the activity greatly enhances the evidence for criticality, revealing it where previous methods would not. The approach is general and
could be used in other biological networks.

\end{abstract}

\maketitle

Complex brain functions usually involve large numbers of
neurons interacting in diverse ways
and spanning a wide range of time and length scales. 
At first sight, systems of inanimate matter seem to enjoy more regular
properties, but they may also display complex and heterogeneous behaviors when in
a critical state, which corresponds to special points of the
parameter space.
Thinking about the brain as a system near a critical
point has been an attractive idea, which has gained attention after the
suggestion that such critical states could be achieved in a
self-organized manner, without fine-tuning \cite{Bak:1988p8276},
but also the proposal that operating near a critical point could be beneficial for
computation \cite{Bertschinger:2004p13235}.

Despite considerable work on the 
foundations of a  theory of critical neural
networks (see \cite{Levina:2007p8058,Magnasco:2009p8347} for recent examples), the validation of these ideas by
experimental data has proven difficult, largely because it requires to measure
the detailed activity of large populations of neurons. Recent progress has
been made possible by the advance of multi-electrode or imaging
techniques, which have helped detect signatures of criticality in a
variety of neural contexts. Two lines of empirical
evidence, rooted in different approaches to
critical systems, have been followed, albeit with little intersection. In line
with the original ideas of self-organised criticality and
branching processes, the statistics
of neural avalanches in cortical layers has been shown to display power-law statistics \cite{Beggs:2003p7980,Petermann:2009p13200,Friedman:2012p12883,Tagliazucchi:2012p13257}. This
observation is indicative of the critical nature of the
system's dynamics, but it relies on arbitrary choices,
such as the number of units considered, the minimal silence time to call the end
of an avalanche, or the definition of a neural event itself.
The stability exponents of the neural
dynamics, which become positive at the 
transition to chaos, have also been used as signatures of criticality \cite{Solovey:2012p13113}. This criterion relies on a
continuous description of neural activity, which is inappropriate for codes relying on combinations of spikes and silences. Both these approaches address the
dynamical aspect of criticality. They
require the definition of an {\em ad hoc} order parameter (avalanche size, firing rates),
which may not be the most relevant one for neural activity. A second line of
enquiry, which focuses on the thermodynamic aspect of criticality, has been to study the frequency of combinations of
spikes and silences in a neural population as a statistical mechanics problem, and explore
its properties in the thermodynamic limit
\cite{Schneidman:2006p1273,Tkacik:2013p13254}, using non-parametric signatures such as the divergence of
the specific heat to demonstrate critical behaviour \cite{Tkacik2006,Tkacik2014}. These analyses
have however been restricted to the simultaneous distribution of neural activity, with no regard to its dynamical properties, which may be
strongly out of equilibrium and may contain important clues about critical behavior. Because of their respective limitations,
neither of these approaches gives us a coherent picture for assessing and understanding all aspects of criticality.

In this paper we overcome these limitations by introducing a framework for analysing the critical
dynamics of neural networks. We apply a thermodynamic approach to
the population's spiking activity over long periods, 
treating time as an extra dimension.
We propose a generalized, time-normalized specific heat of spike
trains as an indicator of critical dynamics.
The approach accounts for the combinatorial nature of the code, and does
not rely on the choice of an order parameter. It reduces to the usual
notion of dynamical criticality through the stability exponents of the dynamics when the
number of spikes can be approximated as a continuous variable. It is also equivalent to the
thermodynamic criticality of \cite{Tkacik:2013p13254,Tkacik2014} when time correlations are
ignored. 
We apply our criterion to a dense population of ganglion cells recorded in the rat retina.
We will show that the dynamics of this population are close to a critical point, where the specific heat diverges. This divergence appears to be much more pronounced once the temporal dynamics are taken into account.

\begin{figure}
\begin{center}
\noindent\includegraphics[width=.49\linewidth]{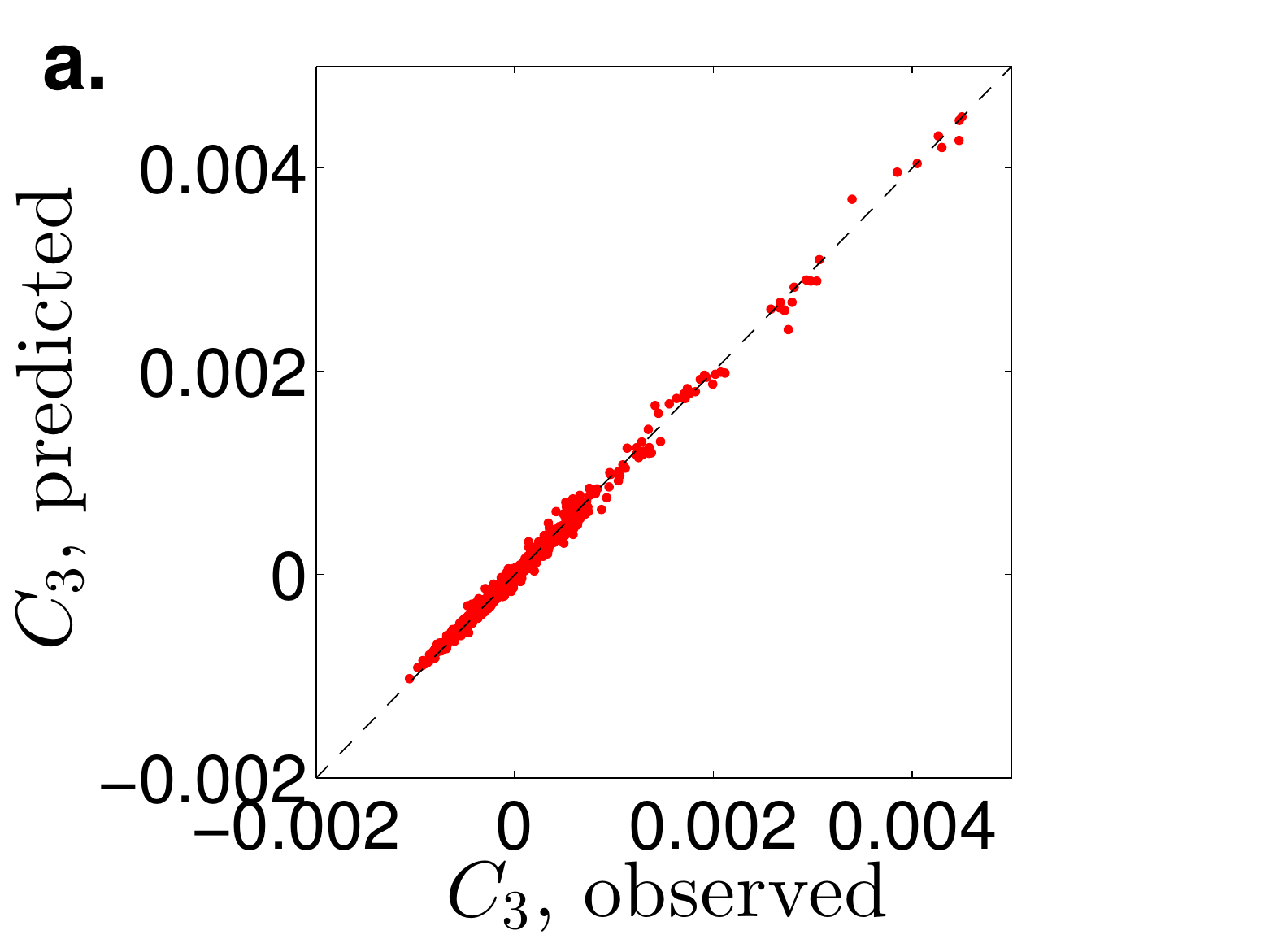}
\noindent\includegraphics[width=.49\linewidth]{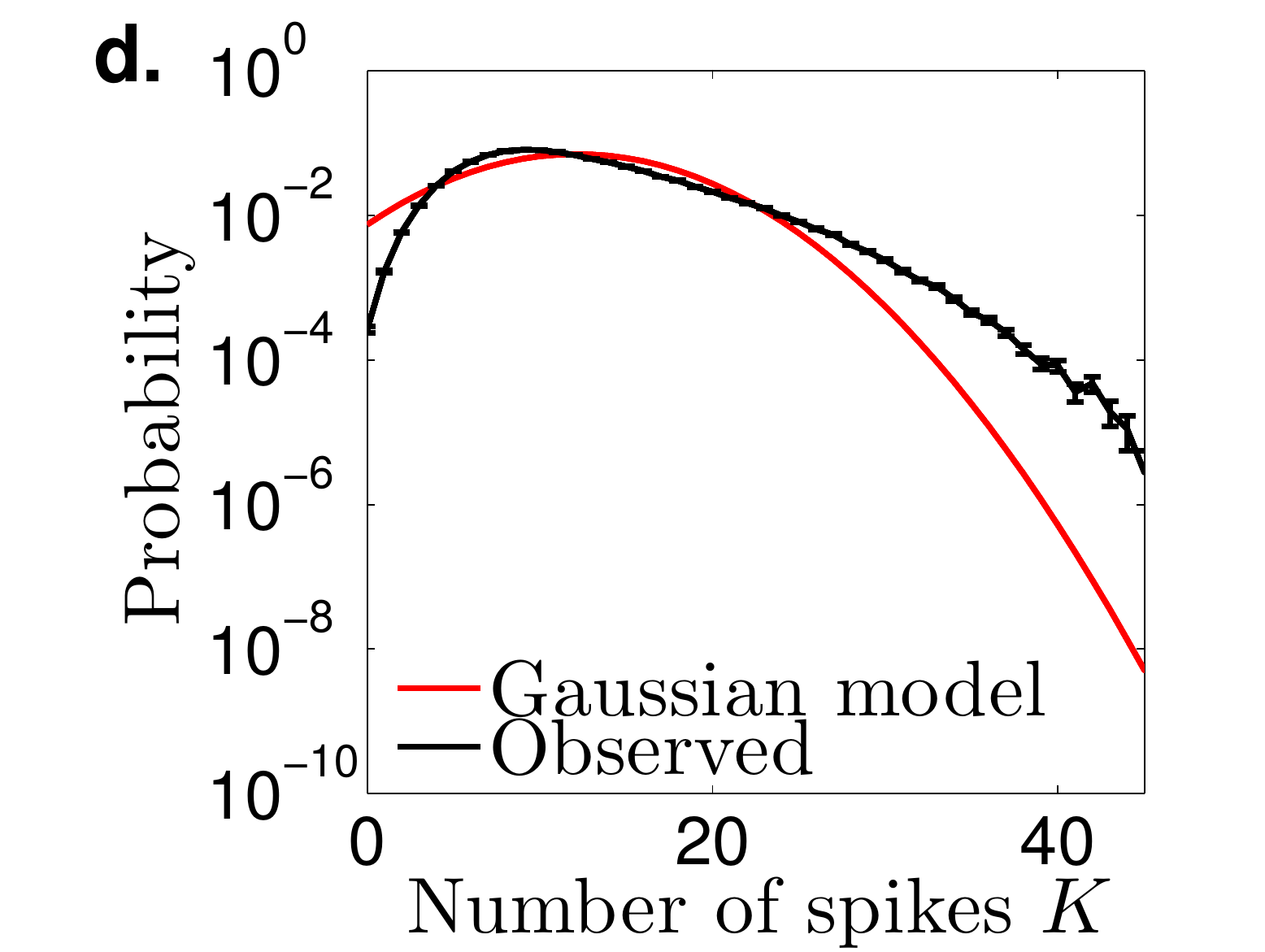}
\noindent\includegraphics[width=.49\linewidth]{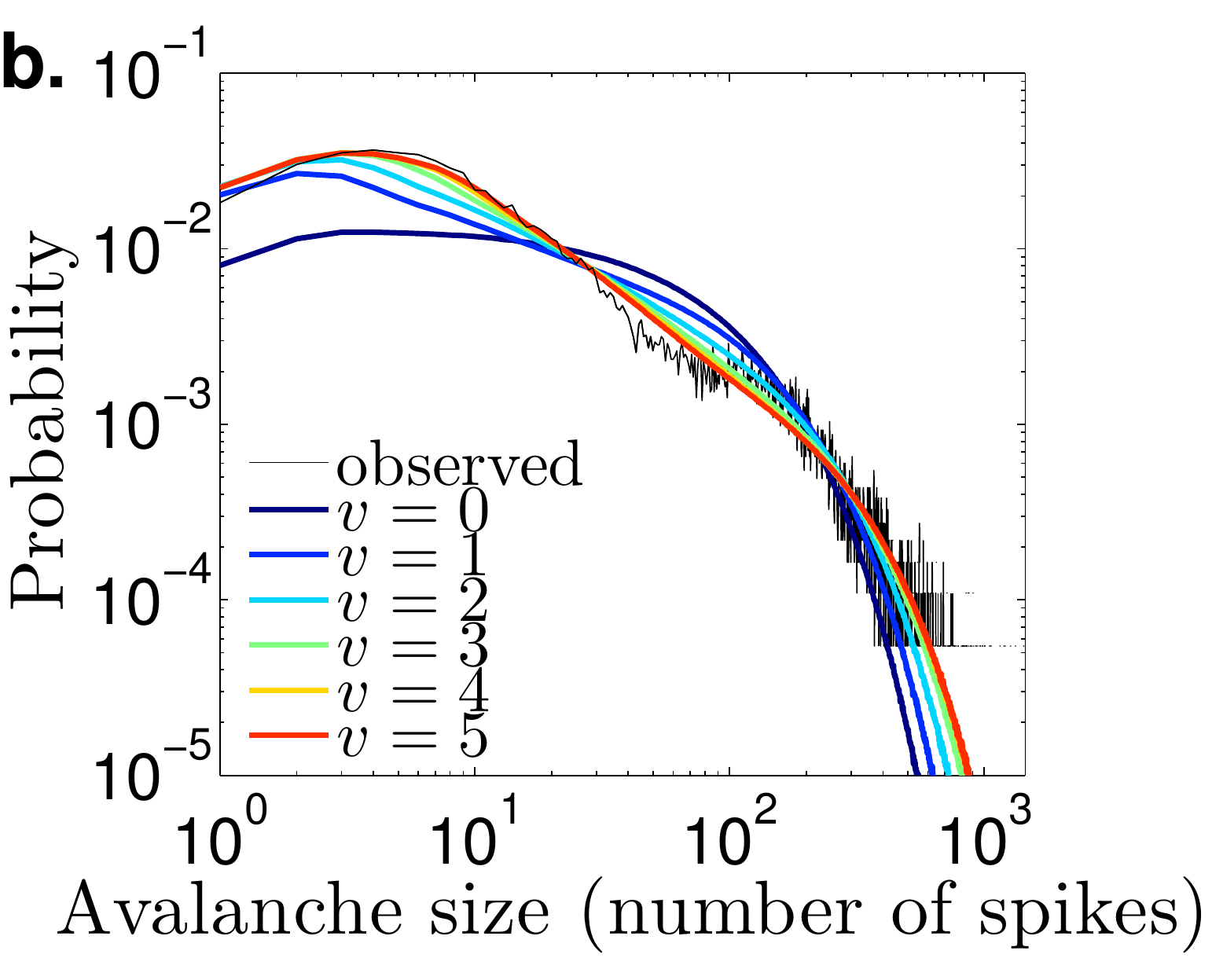}
\noindent\includegraphics[width=.49\linewidth]{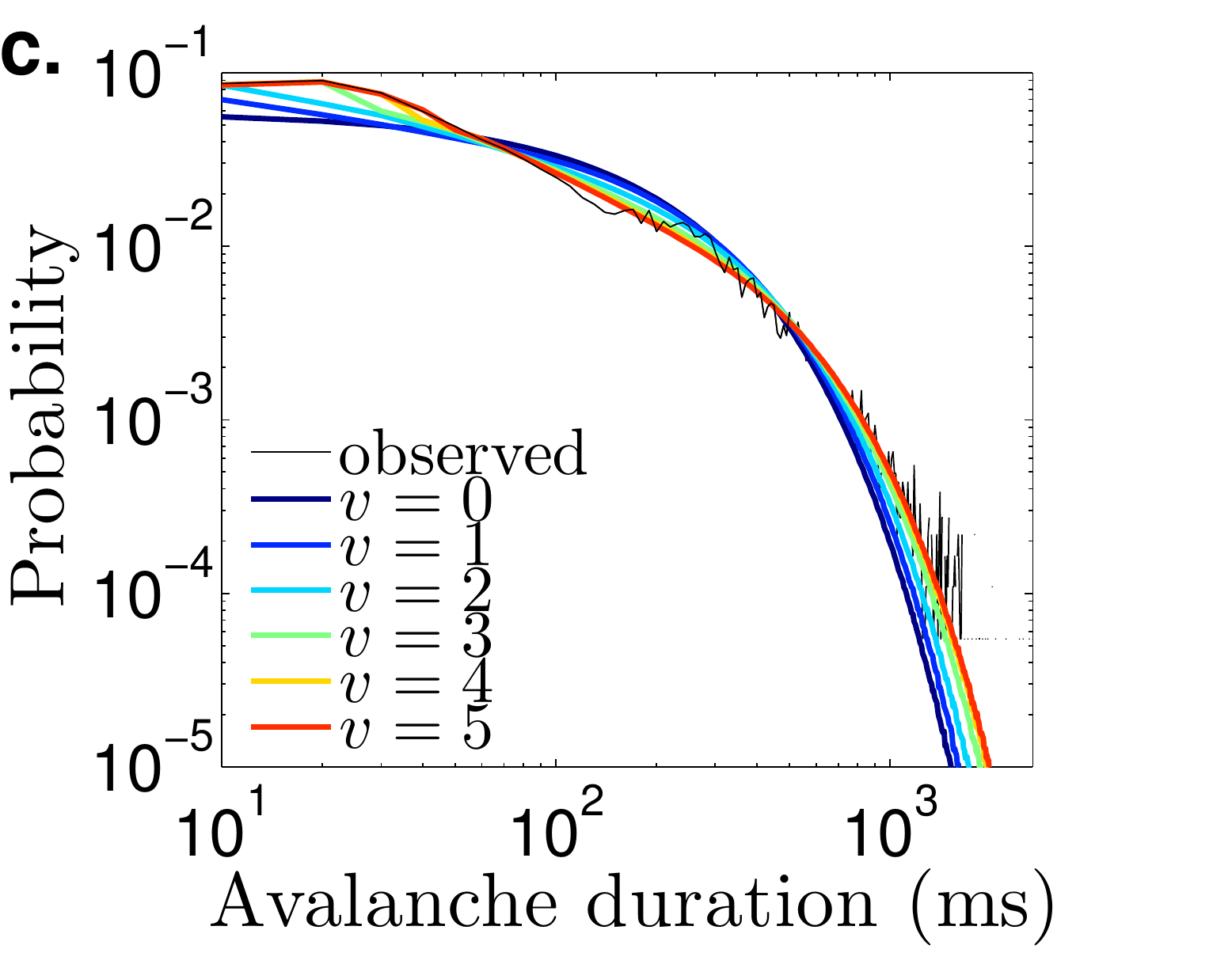}
\caption{{\bf The model captures the global dynamics of the network.}
(a) Predicted versus observed connected correlation functions $C_3=P(K_{t},K_{t+1},K_{t+2})-P(K_{t+2})P(K_{t+1})P(K_t)$ between the total number of spiking neurons in three consecutive time windows of length $\Delta t=10$ms, for a subnetwork of $N=61$ neurons. 
(b) and (c) Model prediction for the size and duration of avalanches, with different temporal ranges $v$, for the same subnetwork of $N=61$ neurons. An avalanche is a series of non-silent 10 ms windows, ended by a silent window. While a model of independent spikewords ($v=0$) is a poor predictor of avalanche statistics, including time correlations over a few time windows greatly improves the prediction. (d) The distribution of the number of spiking neurons in a window $\Delta t=10$ms (black curve) is exactly fitted by the model, by construction. By contrast, it is not well predicted by a Gaussian model (red curve).
\label{fig:tests}
}
\end{center}
\end{figure}

To describe the discrete spiking activity of a population of $N$
neurons, we divide time into small windows of length $\Delta t$, and
assign a binary variable $\sigma_{i;t}=1$ if neuron $i$ has spiked at
least once within window $t$, and $0$ otherwise. $\Delta t$ must be small enough so that two spikes are unlikely to occur in the same window. In the following we will take $\Delta t=10$ms.
The probability of a given
multi-neuron spike train between $t=1$ and $t=L$, or generalized
``codeword'' $\{\sigma_{i,t}\}$, can formally be written in a
Boltzman form:
\begin{equation}\label{eq:boltzmann}
P_{\beta}(\{\sigma_{i,t}\})=\frac{1}{Z(\beta)}e^{-\beta E(\{\sigma_{i,t}\})},
\end{equation}
where $Z(\beta)$ is a normalization constant. By analogy to equilibrium statistical mechanics, $E$ is interpreted as the energy of the spike train. In information-theoretic terms, the surprise of the spike train is related to its energy through $-\log P=\beta E+\log Z(\beta)$.
$\beta$ is
an adjustable control parameter equivalent to an inverse temperature,
set to 1 by convention to describe the observed spike
statistics. Its function is to study the parameter space of
models in the vicinity of the actual system at $\beta=1$, and thus assess its proximity to a critical state.

One possible indicator for detecting a critical point is the specific heat \cite{Chaikin,Mora:2011p12729}, defined
in our formalism as:
\begin{equation}
c(\beta)=
\frac{1}{NT} \beta\frac{\partial}{\partial
  \beta}\beta^2\frac{\partial}{\partial \beta} \frac{\log
  Z(\beta)}{\beta} = 
\frac{\beta^2}{NL}\<\delta E^2\>_\beta,
\end{equation}
where $\delta E=E-\<E\>$ denotes fluctuations from the mean energy, and $\<\cdot\>_\beta$ denotes averages taken under probability law $P_\beta$ (see Appendix B). The specific heat has a clear biological interpretation in terms of the spike train statistics:
it is the normalized variance
of the surprise of neural spike trains, $\mathrm{Var}(\log P)/NL$. It quantifies the breadth
 of codeword utilization: $c(\beta)=0$ means that all utilized codewords have uniform usage probability, whereas a large $c(\beta)$ means that the code is balanced between a few frequent codewords and many more rare codewords \cite{Mora:2011p12729}.
We included the normalization $NL$ because the variance of the surprise is expected to be an extensive quantity scaling linearly with the system size, taken both across neurons and time. Thus, in the limit where spiking events $\sigma_{i,t}$ are independent or weakly correlated, $c(\beta)$ should converge to a finite value as $N$ and $L\to\infty$. For example, if all spiking events were independent with the same spiking probabibility $p$ in each time window, we would have $c(\beta) = \beta^2 (pq)^\beta (\log p-\log q)^2/(p^\beta+q^\beta)$ with $q=1-p$,  for all $N$ and $L$ (see Appendix B).
However, if the system is strongly interacting (between neurons, across time, or both) the specific heat may diverge for a certain critical value of the control parameters.
Treating time windows and neurons on equal footing allows us to address both the many-body nature of the problem and its critical dynamics with a single criterion. Since this criterion is based on the surprise, which follows directly from the probabilistic nature of the process, it does not require us to choose an order parameter (spike rates, size of avalanche, {\em etc.}).

Using the divergence of this specific heat as a diagnostic tool for criticality generalizes previous approaches. Firstly, in
the limit $L=1$, where codewords are simultaneous combinations of
neurons and silences, with no regards to the dynamics, we recover the static thermodynamic approach of
\cite{Tkacik2014}. Secondly, the method is consistent with the notion of dynamical criticality based on stability exponents.
Let us assume that the dynamics is well
described by a single projection of the spikes onto a continuous variable, {\em e.g.}
$K_t=\sum_{i=1}^N \sigma_{i,t}$, and linearized to a Gaussian, Markovian dynamics
\beq\label{eq:autoreg}
K_{t+1}=a K_t + b+\textrm{Gaussian noise},
\eeq
where the
stability exponent of the dynamics is $\log(a)<0$.
This system is critical for $a\sim 1$; above the transition, the linearized dynamics breaks down as the system becomes chaotic.
The specific heat of this model
at $\beta=1$,
\begin{equation}\label{eq:specheat}
c(\beta=1)\sim {\left(\log\frac{N-\<K\>}{\<K\>}\right)}^2 \frac{\mathrm{Var}(K)}{N}
\frac{1+a}{1-a},
\end{equation}
diverges at the critical point $a=1$ (see Appendix C).
Lastly, the approach can detect criticality in simple models of neural avalanches.
Consider the spiking model proposed in \cite{Beggs:2003p7980}, where a neuron $i$ spikes at time $t+1$ in response to a pre-synaptic neuron $j$ spiking at time $t$ with probability $p_{ij}$ (Appendix D). This model is parametrized by the branching parameter $\omega=(1/N)\sum_{ij}p_{ij}$, which controls the spread of neural avalanches. At the critical point $\omega=1$ the system exhibits avalanches with power-law statistics. We estimated the specific heat $c(\beta=1)$ of that model numerically, and found it to diverge with the system size precisely at the critical value of the branching parameter $\omega=1$ (Fig.~S1). In sum, the specific heat, when defined on the temporal statistics of spike trains, allows us to detect dynamical critical transitions, without having to know the order parameter or the definition of an avalanche.

Our goal is to apply our criterion to the spiking activity of a dense population of $N=185$ retinal ganglion cells in the rat retina \cite{Marre:2012p12803}, stimulated by films of randomly moving bars and binned with $\Delta t=10$ms (see Methods). However, to carry out our analysis we first need to learn a  probabilistic law $P(\{\sigma_{i;t}\})$ from the spike trains, which in itself can be a daunting task. We do so by employing the principle of maximum entropy \cite{Jaynes:1957p4009,Jaynes:1957p4011,Schneidman:2006p1273}, which consists in finding the least constrained distribution of spike trains ({\em i.e.} of maximum entropy  $-\sum P\log P$) consistent with a few selected observables of the data (see Appendix E).
In \cite{Tkacik:2013p13254} the global network activity of the salamander retina was modeled by constraining the distribution 
$P(K)$ of the total number of spikes in the population
(see also \cite{Okun:2012p12844}). The inferred model was shown to be near a critical point.
However that choice of constraints did not address the dynamical nature of the spike trains. To do that while making as few additionnal assumptions as possible, we also constrain a dynamical quantity --\,the joint distribution of $K_t$ at two different times $P_u(K_t,K_{t+u})$. This leads to a family of time translation invariant models of the form in Eq.~\ref{eq:boltzmann} with:
\beq\label{eq:energy}
E=-\sum_t h(K_t)-\sum_{t}\sum_{u=1}^v J_u(K_t,K_{t+u}),
\eeq
where $v$ is the model's temporal range --\,the larger the $v$, the more accurate the model.
Applying the maximum entropy principle to trajectories rather than instantaneous states is sometimes also refered to as the maximum caliber method \cite{Presse:2013p13122}.
The model is learned by fitting the parameters $h(K)$ and $J_u(K,K')$ to the data using the technique of transfer matrices (see Appendices G and H). We find that a temporal range of $v=4$ suffices to account for the temporal correlations of $K$ (see Appendix I and Figs.~S2 and S3).

The obtained model reproduces key dynamical features of the data. Fig.~\ref{fig:tests}a compares data and model for the joint distribution of the numbers of spikes in three consecutive time windows, showing excellent agreement despite this observable not being fitted by the model. More importantly, the model predicts well the distributions of size and duration of neural avalanches, defined as continuous epochs of $K>0$, as shown in Fig.~\ref{fig:tests}b and c for a subset of $N=61$ neurons. The agreement extends over seconds, way beyond the model's temporal range of $v\times 10$ms $=50$ms. Although we will not use avalanche statistics to discuss criticality in this paper, as is often done \cite{Beggs:2003p7980,Petermann:2009p13200,Friedman:2012p12883,Tagliazucchi:2012p13257}, the success of our model in predicting them demonstrates its ability to capture complex, collective dynamical behaviour.

Simplifying the model further does not capture important statistics of the data.
We could make a continuous approximation for $K_t$ and constrain only the first two moments of the distributions. This approximation would yield an autoregressive model generalizing Eq.~\ref{eq:autoreg} (see Appendix F). However, the statistics of such models would all be Gaussian, in plain contradition with the observed distribution of spikes $P(K)$, see Fig.~\ref{fig:tests}d. Since the tail of that distribution is related to the collective properties of the population \cite{Tkacik:2013p13254}, it is important to account for it fully, and our model is the simplest one that does that.

\begin{figure}
\begin{center}
\noindent\includegraphics[width=.79\linewidth]{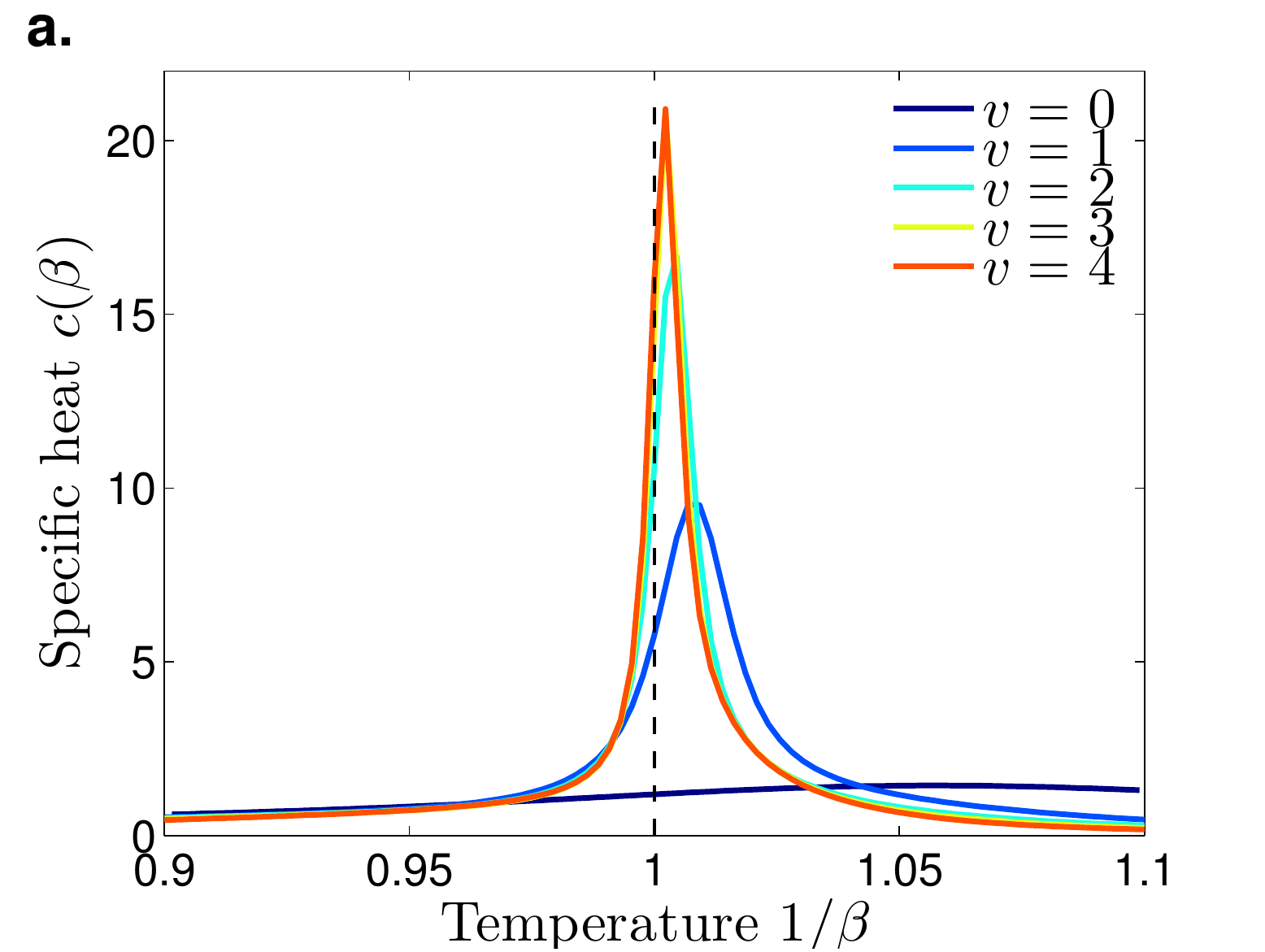}
\noindent\includegraphics[width=.79\linewidth]{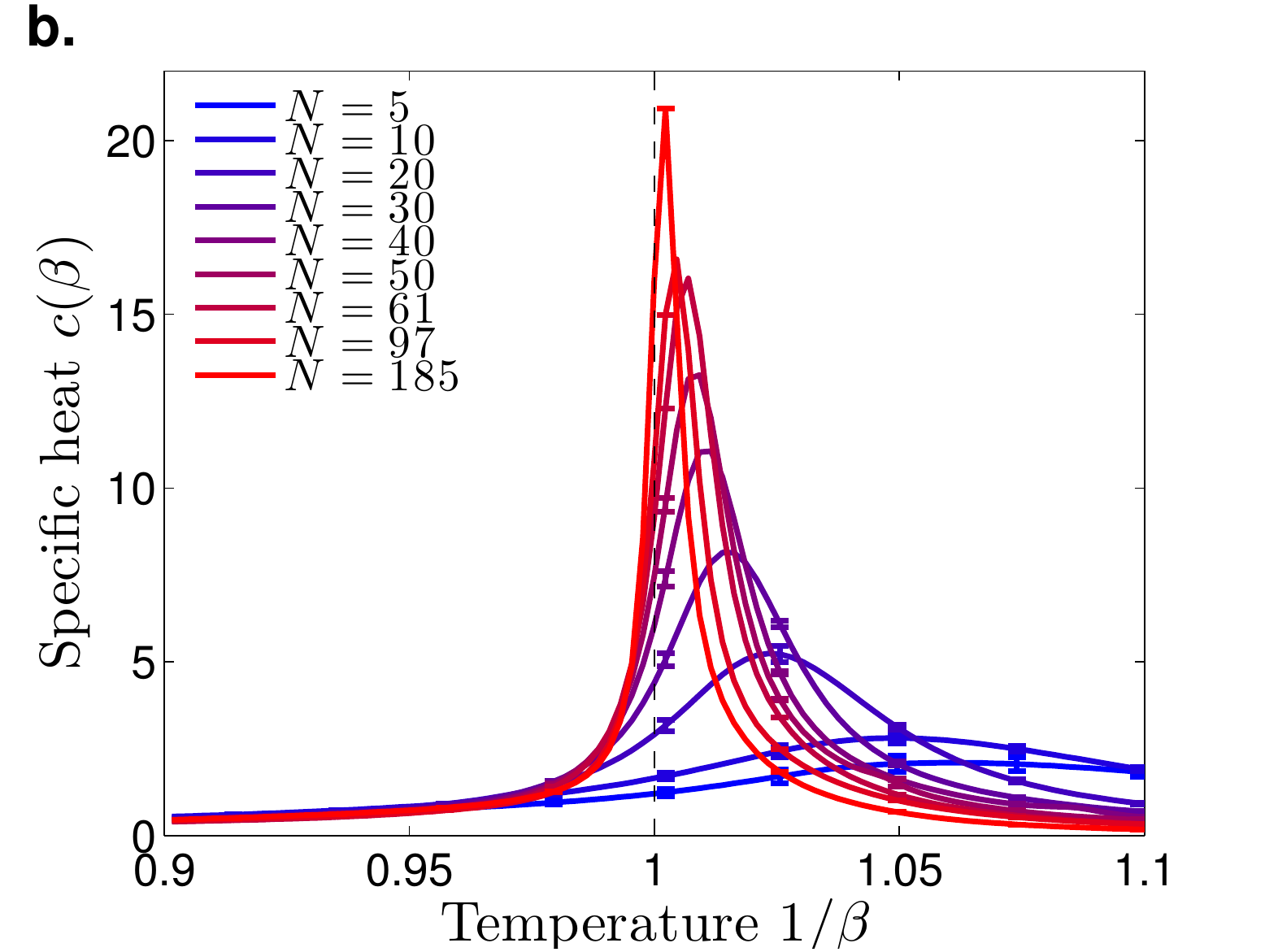}
\caption{{\bf Divergence of the specific heat of spike trains.} (a)
Specific heat $c(\beta)$ of spike trains of the entire population ($N=185$), as a function of the temperature $1/\beta$, for an increasing temporal range $v$. Temperature $\beta=1$ corresponds to the observed statistics of spike trains. The curve with $v=4$, which fully accounts for the dynamics of the spike trains, shows a markedly higher peak than that obtained from the statistics of instantaneous codewords ($v=0$). (b) Specific heat of spike trains for subnetworks of increasing sizes $N$, for $v=4$. Each point is averaged over 100 random subnetworks for $N\leq 50$, and shows one representative network for $N=61$ and $97$. The error bars show standard deviations. The peak increases with network size, indicating a divergence in the thermodynamic limit.
\label{fig:specheat}
}
\end{center}
\end{figure}

\begin{figure}
\begin{center}
\noindent\includegraphics[width=.49\linewidth]{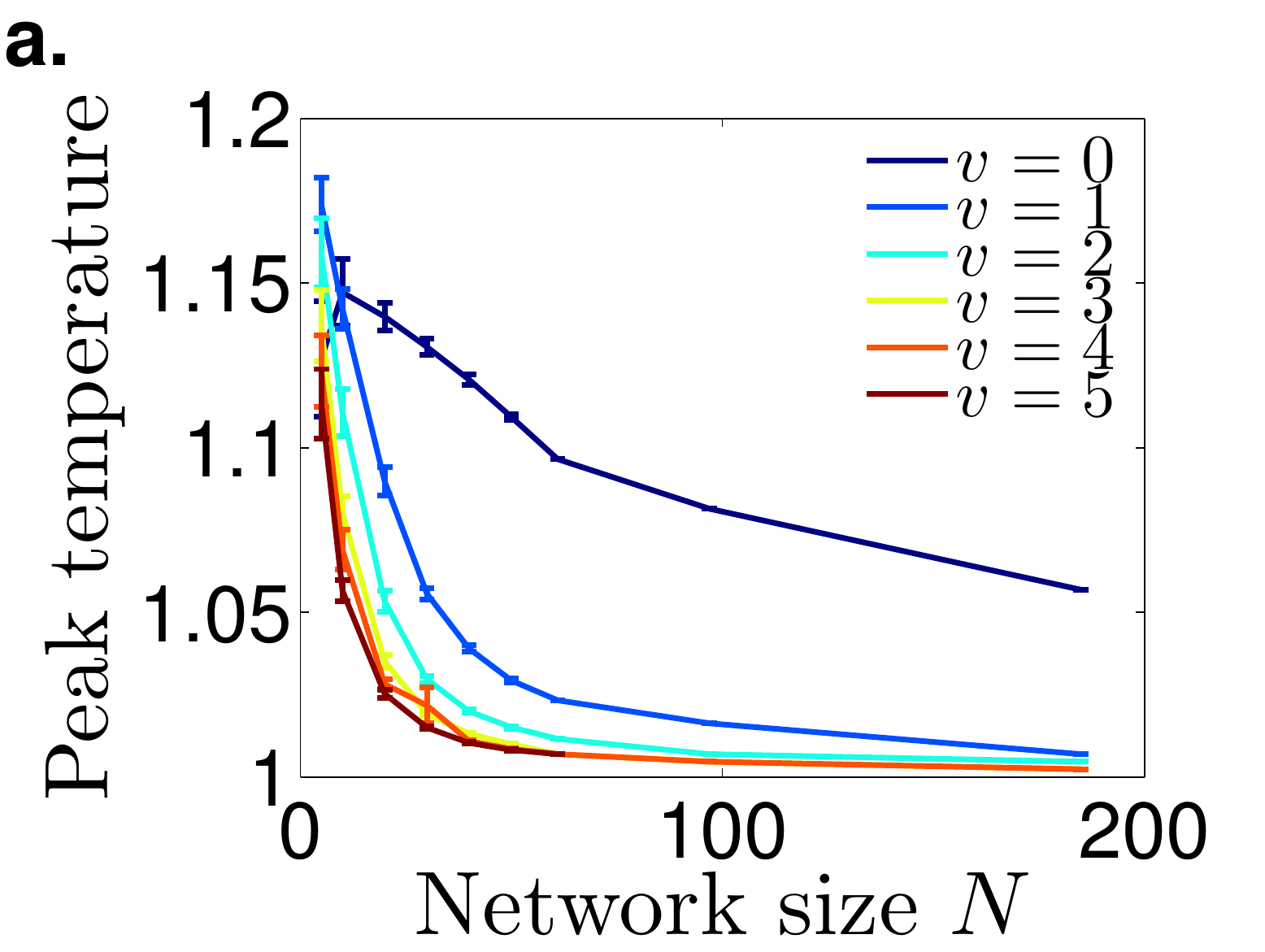}
\noindent\includegraphics[width=.49\linewidth]{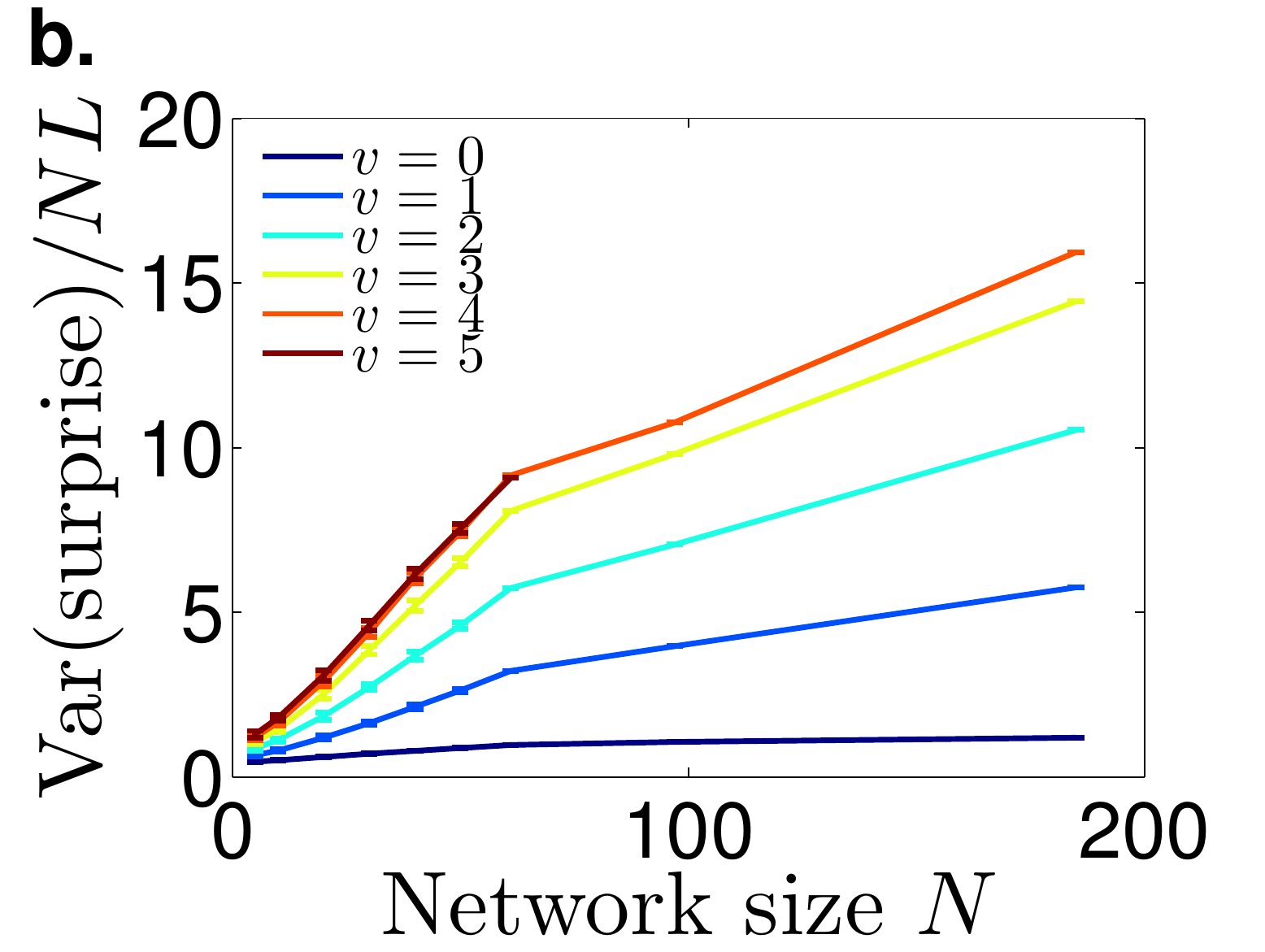}
\caption{{\bf Finite-size scaling.} (a) Position of the peak  $1/\beta$ in specific heat (see Fig.~\ref{fig:specheat}) as a function of network size $N$, for increasing time ranges $v$. Accounting for the dynanics of spike trains ($v>0$) gives peaks that are much closer to the temperature of real spike trains ($\beta=1$) than for instantaneous spikewords ($v=0$). (b) The specific heat of real spike trains, $c(\beta=1)$, is equal to the variance of the surprise per neuron and per unit time, $\mathrm{Var}(\log P)/NL$. This variance increases with the system size $N$ and with the temporal range $v$. Note that the $v=5$ curves are very close to the $v=4$ ones up to $N=61$ (above which they are not calculated).
\label{fig:surprise}
}
\end{center}
\end{figure}

Confident that our model gives a precise account of the temporal dynamics of the global network activity, we can use it to estimate its specific heat.
\fig{fig:specheat}A represents the specific heat of the learned models (Eq.~\ref{eq:energy}) for all $N=185$ neurons as a function of the temperature $1/\beta$, for different choices of the temporal range $v$. The special case $v=0$, in which time correlations are ignored, shows only a moderate peak in specific heat, and far from $\beta=1$. By contrast, including time correlations ($v>0$) greatly enhances the peak, which rapidly approaches $\beta=1$ as the temporal range $v$ is increased. Fig.~\ref{fig:specheat}B shows how the peak in specific heat behaves for random subgroups of neurons of increasing size, for $v=4$. Similarly, the peak becomes larger, sharper and closer to  $\beta=1$ as the network size grows. These are striking results, if we recall that all these curves would fall on top of each other for independent (or weakly correlated) spiking events.
The unusal scalings of Fig.~\ref{fig:specheat} suggest that the system is indeed close to a critical point. But they also show that both the collective behaviour of the population and the temporal correlations play a crucial role in revealing the critical properties of the network. In fact, the convergence of the peak of the specific heat towards $\beta=1$ is only apparent when time correlations are taken into account ($v>0$), as illustrated by Fig.~\ref{fig:surprise}A. This is in contrast with the results of \cite{Tkacik:2013p13254}, which found signatures of criticality even for $v=0$, although this apparent disagreement may be attributed to differences between species (the rat having much higher average firing rates in their retinal ganglion cells than the salamander). 

Although the peak of the specific heat is a somewhat abstract quantity, the fact that it increases and approaches $\beta\sim 1$ implies that  the normalized variance of the surprise $c(\beta=1)=\mathrm{Var}(\log P)/NL$ also increases with the system size, as shown in Fig.~\ref{fig:surprise}B. These variances are extremely high compared to that we would obtain if all spiking events were independent, $c_{\rm inde}(\beta=1)=0.38$, indicating a very wide spectrum of codeword usage.

For the sake of simplicity and tractability,
we have here only modelled the global network activity of the retina.
Although these models capture important features of the dynamics (Fig.~\ref{fig:tests}), more detailed models accounting for the full temporal cross-correlations between individual neurons \cite{Marre:2009p13087,Vasquez:2012p13088} could provide us with a more precise description of the spiking dynamics, and better approximations to the specific heat curves. 
In principle our inference procedure may also depend on the choice of window size $\Delta t$.
We repeated the analysis for windows of $5$ms, and found the same results, with an excellent agreement between models that have a different $\Delta t$ but the same temporal range $v\times \Delta t$ expressed in seconds (see Appendix J and Figs.~S4 and S5).

We have introduced a framework for studying the collective dynamics of a population of neurons. This formalism provides us with a non-parametric criterion for detecting the proximity to a critical state, whether this criticality stems from strong collective effects in the population, from critical dynamics at the edge of chaos, or from both, thus generalizing previous approaches. 
When we apply our approach to large-scale recordings in the retina, we find that the population dynamics are very close to a critical state.
Compared to the static thermodynamic approach of \cite{Tkacik:2013p13254,Tkacik2014}, which focused on the statistics of instantaneous codewords, the peak in specific heat that we find is 10 times larger, and much closer to the system's actual temperature of $1$. Our results suggest that although simultaneous correlations between neurons are an important marker of near-critical behavior, accounting for their dynamical component greatly enhances our confidence and understanding of it.

The idea that biological systems may operate near a critical point is not restricted to the case of neurons \cite{Mora:2011p12729}, with evidence in systems as diverse as the cochlea \cite{Eguiluz:2000p11147}, immune repertoires \cite{Mora:2010p5398}, natural images \cite{Stephens:2013p13259}, animal flocks \cite{Bialek:2014p13205,Attanasi:2013p13048,Cavagna:2014p13203} or the regulation of genes in early fly developpment \cite{Krotov:2014p13258}, to name but just a few, and we expect our approach to be useful when both the collective behaviour and the dynamics play an important role.

We thank William Bialek and Gasper Tkacik for helpful comments on the manuscript.
This work was supported by the DEFI-SENS 2013 program from the Centre National de la Recherche Scientifique, by the Agence Nationale pour la Recherche (ANR: OPTIMA), and by the French State program ``Investissements d'Avenir'' managed by the Agence Nationale de la Recherche [LIFESENSES: ANR-10-LABX-65].
SD was supported by a PhD fellowship from the region Ile-de-France.

\section*{Methods}

\paragraph*{Retinal recordings.}
Recordings were performed on the Long-Evans adult rat. In brief, animals were euthanized according to institutional animal care
standards. The retina was isolated from the eye under dim
illumination and transferred as quickly as possible into oxygenated
AMES medium. The retina was then lowered with
the ganglion cell side against a multi-electrode array whose electrodes were spaced by 60 microns, as previously described \cite{Marre:2012p12803}. Raw voltage traces were
digitized and stored for off-line analysis using a 252-channel
preamplifier (MultiChannel Systems, Germany). The recordings
were sorted using custom spike sorting software developed
specifically for these arrays \cite{Marre:2012p12803}. We extracted the activity from 185 neurons with satisfying standard tests of stability and limited number of refractory period violations. 
\paragraph*{Visual stimulation.}
Our stimulus was composed of several black bars moving randomly on a gray background. The trajectory was a random walk with a restoring force to keep the bar close to the array (see Appendix A for details). The stimulus was displayed using a Digital Mirror Device and focused on the photoreceptor plane using standard optics. The analysed data corresponds to one hour ($L=360,000$ with $\Delta t=10$ms) of recordings.

\bibliographystyle{pnas}
\bibliography{thierry,byhand}

\appendix

\setcounter{figure}{0}
\makeatletter 
\renewcommand{\thefigure}{S\@arabic\c@figure}
\makeatother

\section{Visual stimulation}
Our stimulus was composed of two black bars moving randomly on a gray background. Each bar was animated by a brownian motion, with additional feedback force to stay above the array, and repulsive forces so that they do not overlap. The bars stay within an area that covers the whole recording array. The amplitude of the bar trajectories allowed them to sweep the whole recording zone. The trajectories of the bars $x_1$ and $x_2$ are described by the following equations:
\begin{eqnarray}
\frac{dv_1}{dt} &=& -\frac{v_1}{\tau} + \mathrm{sign}(x_1 - x_2){\left(\frac{R}{|x_1 - x_2|}\right)}^6 \nonumber\\
&& - \omega_0^2 (x_1 - \mu_1) + \sigma\, W_1(t)\\
\frac{dv_2}{dt} &=& -\frac{v_2}{\tau}  + \mathrm{sign}(x_2 - x_1){\left(\frac{R}{|x_2 - x_1|}\right)}^6\nonumber\\
&& - \omega_0^2 (x_2 - \mu_2) + \sigma\, W_2(t) 
\end{eqnarray}
where $W_1(t)$ and $W_2(t)$ are two Gaussian white noises of unit amplitude,  $\mu_2- \mu_1= 600 \mu m$ is the shift between the means, $\omega_0=1.04$\,Hz and $\tau=16.7$\,ms. The width of one bar is $100 \mu m$. The stimulus was displayed using a Digital Mirror Device and focused on the photoreceptor plane using standard optics.

\section{Thermodynamics of spike trains}
Let us start with the probability distribution for entire spike trains $\{\sigma_{i,t}\}$, $i=1,\ldots,N$, $t=1,\ldots,L$. By analogy with Boltzmann law we can write this probability as: 
\beq\label{eq:boltzmannT1}
P(\{\sigma_{i,t}\})=\frac{1}{Z}e^{-E(\{\sigma_{i,t}\})},
\eeq
where $E(\{\sigma_{i,t}\})$ and $-\log Z$ are defined up to a common constant. The surprise $-\log P(\{\sigma_{i,t}\})$ is equal to $E(\{\sigma_{i,t}\})+\log Z$.
Note that considering the statistics of entire spike trains over time allows for a well-defined $\Delta t\to 0$ limit, with the concomitant scaling $L \sim 1/\Delta t$, by contrast to the static thermodynamic approach ($L=1$) where this limit tends to the all-silent state with probability one.

The probability distribution in Eq.~\ref{eq:boltzmannT1} will produce typical spike trains with the same statistics as the experiment. To explore this model across a line in parameter space, we can generalize Eq.~\ref{eq:boltzmannT1} to an arbitrary fictious temperature:
\beq\label{eq:boltzmannT}
P_\beta(\{\sigma_{i,t}\})=\frac{1}{Z(\beta)}e^{-\beta E(\{\sigma_{i,t}\})}.
\eeq
While $P_{\beta=1}$ describes ``typical'' spike trains with the same statistics as the experiment, this generalized distribution allows us to explore atypical spike trains of low or high energy (accessed by high and low $\beta$), or equivalently of high and low surprises.

The free energy is defined as $F(\beta)=-\beta^{-1}\log Z(\beta)$. The Shannon entropy of $P_{\beta}$,
\beq
S(\beta)=-\sum_{\{\sigma_{i,t}\}} P_\beta(\{\sigma_{i,t}\})\log P_{\beta}(\{\sigma_{i,t}\}),
\eeq
can be calculated as $S(\beta)=\partial F/\partial \beta=\beta^{-1} (\<E\>_\beta -F(\beta))$, where $\<\cdot\>_{\beta}$ denotes an average taken over spike trains with probability law $P_{\beta}$. This last relation is better known in the form $F=E-TS$, with $T=\beta^{-1}$ is  temperature. The heat capacity is defined as:
\beq\label{eq:heatcapacity}
C(\beta)=T\frac{\partial S}{\partial T}=-\beta \frac{\partial S}{\partial \beta}=\beta^2(\<E^2\>_\beta-\<E\>^2_\beta).
\eeq
In statistical physics it is an extensive quantity, meaning that it scales with the system size $NL$. The specific heat $c(\beta)=C(\beta)/NL$ is the heat capacity normalized by the system size.

Let us consider a simple example, where each neuron spikes with probabily $p_i=r_i\Delta t$ in each time window (where $r_i$ is its spike rate), independently of the other neurons and of its own spiking history. In the limit $\Delta t\to 0$ these are just Poisson neurons.
The probability of a given spike train factorizes over neurons and over time, and reads:
\beq
P_{\beta}(\{\sigma_{i,t}\})=\prod_{t=1}^L \prod_{i=1}^N \frac{1}{z_i(\beta)} p_i^{\beta\sigma_{i,t}}q_i^{\beta(1-\sigma_{i,t})},
\eeq
where $q_i=1-p_i$ and $z_i(\beta)=p_i^\beta+q_i^\beta$. The specific heat can be calculated from Eq.~\ref{eq:heatcapacity}:
\beq
c(\beta)=\frac{1}{N} \sum_{i=1}^N \beta^2 (p_iq_i)^\beta (\log p_i-\log q_i)^2/(p_i^\beta+q_i^\beta).
\eeq
This expression has no divergence as a function of $\beta$. 
For small uniform spiking probability $p_i=p\ll 1$, the specific heat
at the natural temperature is also small: $c(\beta=1)\sim p (\log
p)^2$. In that same limit, the peak in specific heat is reached at
high temperatures, $\beta_c \sim -\alpha/\log p$, where $\alpha\approx
2.2$ is solution of the irrational equation $\alpha=2(1+e^{-\alpha})$;
the value of the peak does not depend on $p$, and is $c(\beta_c)\sim \alpha(\alpha-2)\approx 0.48$.

\section{Thermodynamics of a simple auto-regressive model}
We consider a simple case where we assume that the neural population is well described by a continuous parameter describing the total number of spiking neurons in a time window. Let us call $K_t=\sum_{i=1}^N \sigma_{i,t}$ that number. Its mean is $\<K\>=rN$, where $r$ is the average spike rate of each cell per time window. We denote $K_t=rN+x_t$. Assuming that $K$ and $N$ are large, we can treat $x_t$ as a continuous variable, and model it by a simple Markov dynamics, or auto-regressive model:
\beq\label{eq:autoreg1}
x_{t+1}=ax_t+\epsilon_t,
\eeq
with $\epsilon_t$ a Gaussian noise of mean zero and variance $\sigma^2$. $x_t$ is of mean zero, and its auto-correlation function of $x_t$ reads:
\beq
\<x_t x_{t'}\>=\frac{\sigma^2}{1-a^2} a^{|t-t'|}=\<x_t^2\> a^{|t-t'|}\equiv frN a^{|t-t'|},
\eeq
where $f=\mathrm{Var}(K)/\<K\>$ is the Fano factor of the number of spiking neurons. $f=1$ when the distribution of $K$ is Poisson. When $a\to 1$, the system becomes critical in the traditional dynamical sense, with a diverging correlation time $-1/\log(a)$. This is the ``stability parameter'' obtained from an auto-regressive model \cite{Solovey:2012p13113}.

For each $K$, the probability of a given spiking pattern is uniform:
\beq\label{eq:sigmacond}
P(\sigma_1,\ldots,\sigma_N|K)=\frac{\delta(\sum_i \sigma_i,K)}{\binom{N}{K}}.
\eeq
Assuming that the system is stationary at $t=1$, the probability of a whole spike train of duration $L$ is thus given by:
\beq\label{eq:logP}
\begin{split}
&\log P(\{\sigma_{i,t}\})=-\frac{L}{2}\log(2\pi \sigma^2)-\frac{x_1^2}{2frN}-\frac{1}{2}\log(2\pi frN)\\
&\qquad -E_K-E_\sigma,
\end{split}
\eeq
with
\bea
E_K&=&-\frac{1}{2\sigma^2}\sum_{t=1}^L (x_{t+1}-ax_t)^2\\
E_\sigma&=&\sum_{t=1}^{L} \left[\log \Gamma(N+1)-\log \Gamma(rN+x_t+1)\right.\nonumber\\
&&\left.-\log\Gamma((1-r)N-x_t+1)\right].
\eea
where we have replaced $\binom{N}{K}=\binom{N}{rN+x_t}$ by its expression in terms of Gamma functions $\Gamma(x)$. The term $-E_K$, combined with the first term on the right-hand side of Eq.~\ref{eq:logP}, corresponds to the Gaussian distribution of $\epsilon_t$ with replacement using Eq.~\ref{eq:autoreg1}. The term $-E_\sigma$ corresponds the conditional distribution in Eq.~\ref{eq:sigmacond}. The second and third terms on the right-hand side correspond to the Gaussian distribution of $x_1$, of zero mean and variance $frN=\mathrm{Var}(K)$.

We expand $E_\sigma$ by assuming that $x\ll N$, using Stirling's formula, and obtain at leading order:
\beq
E_\sigma\approx \sum_t NH\left(\frac{Nr+x_t}{N}\right)\approx \sum_t \left[NH(r) +\log\frac{1-r}{r} x_t\right],
\eeq
where $H(x)=-x\log(x)-(1-x)\log(1-x)$ is the binary entropy.

If we neglect terms containing the initial condition $x_1$, the total surprise is, up to a constant, equal to $E_K+E_\sigma$. Its variance, also called heat capacity by analogy with statistical mechanics, is given by
\beq
C(\beta=1)=(\<E_K^2\>-\<E_K\>^2)+(\<E_\sigma^2\>-\<E_\sigma\>^2),
\eeq
as the cross-correlation term involves the third moments of $x_t$ and thus is zero. A calculation using Gaussian integration rules gives, at leading order in the limit $L\to\infty$:
\beq
\<E_K^2\>-\<E_K\>^2=\frac{L}{2}.
\eeq
On the other hand we obtain:
\beq
\<E_\sigma^2\>-\<E_\sigma\>^2=NL {\left(\log\frac{1-r}{r}\right)}^2 fr \frac{1+a}{1-a}.
\eeq
Both variances scale linearly with $L$. This is consistent with the extensivity of the heat capacity: the average surprise scales linearly with $L$, and its variance does as well. But only the second part of the variance scales linearly with $N$. Thus in the limit $N$, $L\to\infty$,
\beq
c(\beta=1)=\frac{C(\beta=1)}{NL}={\left(\log\frac{1-r}{r}\right)}^2 fr \frac{1+a}{1-a}.
\eeq
The variance of the surprise diverges as $a\to 1$, {\em i.e.} as the system becomes critical in the usual dynamical sense. When the Fano factor $f=\mathrm{Var}(K)/\<K\>$ diverges with $N$, the specific heat $c(\beta=1)$ diverges as well. This is the case when fluctuations of $K$ are of the same order of magnitude as $K$ itself, {\em e.g.} $\mathrm{Var}(K)\sim K^2$ and thus $f\sim K\sim rN$, as was observed in the salamander retina \cite{Tkacik:2013p13254}.

\section{Thermodynamics of a model of neural avalanches}
We now study a simple model of spiking dynamics that is known to display critical avalanche statistics \cite{Beggs:2003p7980}. We will show that applying our specific heat criterion allows us to detect the critical point. In this model, neuron $i$ spikes at time $t$ if it receives signal from at least one other neuron $j$, which happens with probability $p_{ij}$, provided that that neuron has spiked at time $t-1$. The probability for a spike train can be written as:
\beq
P(\{\sigma_{i,t}\})=\prod_t \prod_{i=1}^N p_i(t)^{\sigma_{i,t}} [1-p_i(t)]^{1-\sigma_{i,t}}
\eeq
where $p_i(t)=1-\prod_j (1-p_{ij})^{\sigma_{i,t-1}}$ is the probability that neuron $i$ spikes at time $t$. The energy of this process can be easily calculated as $E=-\log(P)=\sum_t \epsilon_t$, with
\beq
\epsilon_t=-\sum_i \sigma_{i,t}\log p_i(t)-(1-\sigma_{i,t})\log[1-p_i(t)]
\eeq

The parameter $\omega=(1/N)\sum_{ij} p_{ij}$ quantifies the probability that a spike generates another spike at the next time step. When $\omega<1$, the spiking activity goes extinct, while when $\omega>1$, it explodes exponentially. Around $\omega\sim 1$, the system is critical and exhibits neural avalanches with power-law statistics \cite{Beggs:2003p7980}. Since the all-silent state is absorbing, in the simulation we further assume that when the system goes into the all-silent state, one random neuron (out of $N$) is made to spike to restart the activity.

\begin{figure}[t]
\begin{center}
\noindent\includegraphics[width=.79\linewidth]{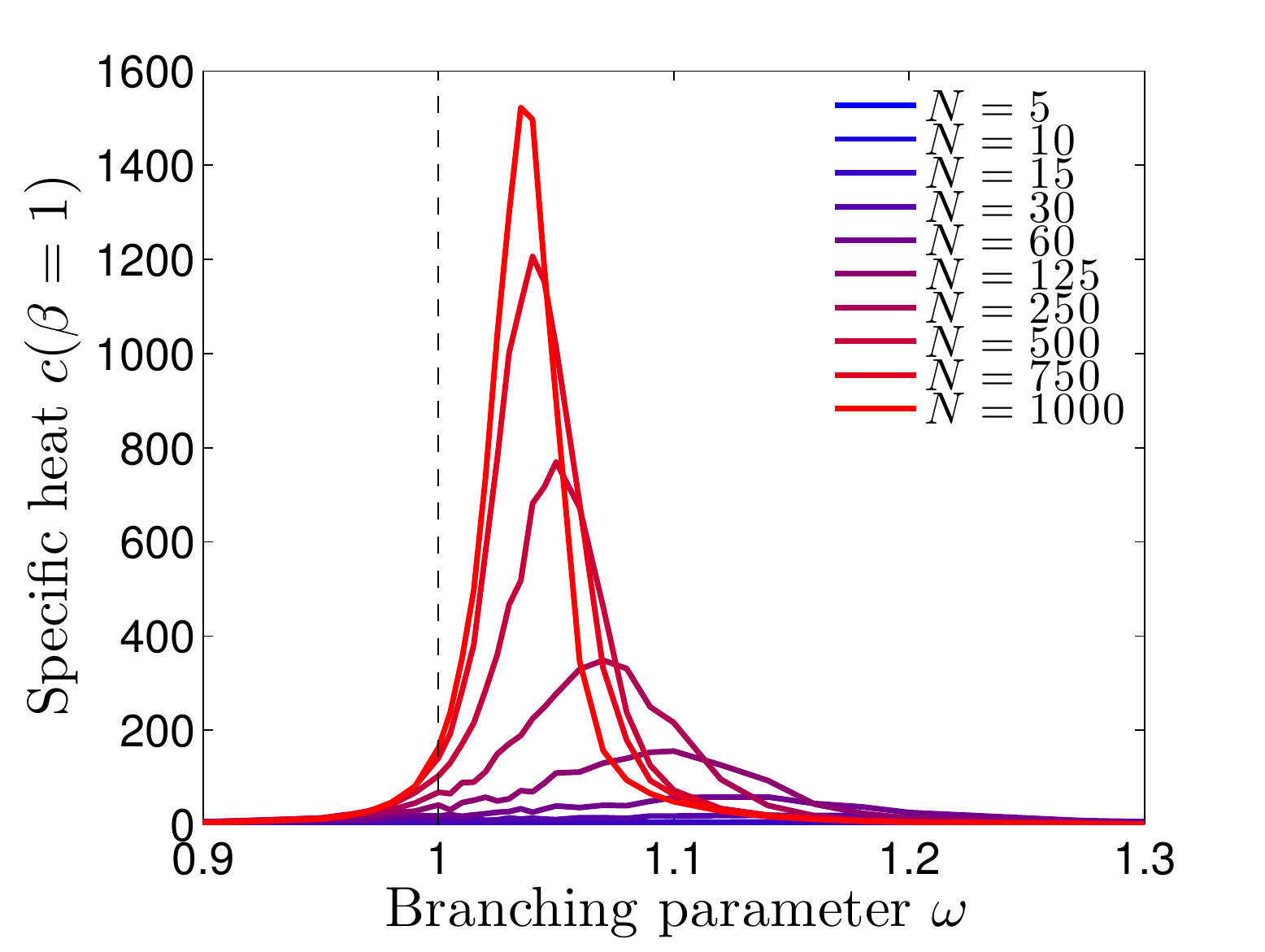}
\caption{{\bf Specific heat of a simple model of neural avalanches.} The specific heat $c(\beta=1)$, or variance of the surprise $\mathrm{Var}(\log P)/NL$, is plotted as a function of the branching parameter $\omega$ in a simple model of neural avalanches, for increasing network sizes $N$. The specific heat gets increasingly peaked as the network size grows, and the peak gets closer to the critical value branching parameter $\omega= 1$.
\label{fig:toy_avalanche}
}
\end{center}
\end{figure}

Taking the $L\to \infty$ limit, the specific heat is just estimated numerically from simulations as 
\beq
c(\beta=1)=\frac{1}{N}\langle\delta \epsilon_t^2\rangle+\frac{2}{N}\sum_{u\geq 1} \langle\delta \epsilon_{t}\delta \epsilon_{t+u}\rangle,
\eeq
where $\delta\epsilon_t=\epsilon_t-\<\epsilon_t\>$.
Fig.~\ref{fig:toy_avalanche} shows the specific heat as function of the branching parameter $\omega$, for increasing network sizes $N$. The specific heat peaks close to $\omega=1$. The peak diverges and gets closer to 1 as the system size is increased. This demonstrates that our criterion for criticality based on the specific heat can help detect a critical transition in this simple model. Note that, in doing so, we have not had to define what an avalanche is. Instead, we have solely relied on the thermodynamic properties of the spike train statistics.

\section{Maximum entropy modeling}
We want to infer a model for the probability of a entire multi-neuron spike train $\{\sigma_{i,t}\}$, $i=1,\ldots,N$, $t=1,\ldots,L$. The principle of maximum entropy allows us to infer an approximation of that probability from measurable observables. We look for a model distribution $P(\{\sigma_{i,t}\})$ that has maximum entropy:
\beq\label{eq:entropy}
-\sum_{\{\sigma_{i,t}\}} P(\{\sigma_{i,t}\})\log P(\{\sigma_{i,t}\})
\eeq
under the constraint that it agrees with the expected value of a few chosen observables $\mathcal{O}_1(\{\sigma_{i,t}\})$, $\mathcal{O}_2(\{\sigma_{i,t}\}),\ldots$, estimated from the data:
\beq\label{eq:constraint}
\<\mathcal{O}_a\>_{\rm data}=\sum_{\{\sigma_{i,t}\}} \mathcal{O}_a (\{\sigma_{i,t}\})P(\{\sigma_{i,t}\}),\ \textrm{for all }a.
\eeq
The technique Lagrange multipliers gives us the form of such a distribution:
\beq\label{eq:maxent}
P(\{\sigma_{i,t}\})=\frac{1}{Z}\exp\left[\sum_a \lambda_a \mathcal{O}_a (\{\sigma_{i,t}\})\right],
\eeq
where $\lambda_a$ are Lagrange multipliers that must be adjusted to
satisfy Eq.~\ref{eq:constraint}, and $Z$ is a normalization constant.

There are many ways to choose the set of observables $\mathcal{O}_a$, and just as many resulting models. Here for simplicity we assume that the system is in a stationary state, so that the statistics of spike trains is time-invariant. This implies that the observables will be time averaged. Our choice of observables are the joint distributions of the number of spiking neurons at different times, $P_u(K_{t},K_{t+u})$, for $u=1,\ldots,v$, defined as:
\beq\label{eq:pairwise}
P_u(K,K')=\frac{1}{L-u} \sum_{t=1}^{L-u} \sum_{\{\sigma_{i,t}\}} \delta_{K,K_{t}}\delta_{K',K_{t+u}} P(\{\sigma_{i,t}\}),
\eeq
where $\delta_{a,b}=1$ if $a=b$ and $0$ otherwise.
The corresponding model of maximum entropy is:
\beq\label{eq:model}
P(\{\sigma_{i,t}\})=\frac{1}{Z}\exp\left[\sum_t h(K_t)+\sum_{t}\sum_{u=1}^v J_u(K_t,K_{t+u})\right]
\eeq
where $h(K)$ and $J_u(K,K')$ are the Lagrange multipliers $\lambda_a$
associated with the constraints on $P_u(K,K')$. Introducing $h(K)$ is
not necessary, because $J_u(K,K')$ suffices to enforce the constraints
on the marginals, but doing so allows us to formally separate first-order from
second-order terms, at the cost of redundancy. As a result, the
definition of the model in Eq.~\ref{eq:model} allows for some freedom
in the definition of the parameters. Indeed the distribution is unchanged upon the transformations:
\beq
J_u(K,K')\to J_u(K,K')+\epsilon(K),\quad h(K)\to h(K)-\epsilon(K),
\eeq
and likewise for the second argument of $J_u$. This degeneracy can be lifted by imposing the following relations:
\bea
\sum_K P(K)h(K)&=&0,\label{eq:C1}\\
\sum_{K} P(K)J_u(K,K')&=&0\ \textrm{for all }K',\label{eq:C2}\\
\sum_{K'} P(K')J_u(K,K')&=&0\ \textrm{for all }K.\label{eq:C3}
\eea
Note that this choice of parametrization does not affect the model distribution itself. It is merely a choice of convention, which ensures that the energy terms $h$ and $J$ are balanced around 0.

In practice it is enough to study the model for $(K_1,\ldots,K_L)$, the distribution of which is:
\beq\label{eq:Kmodel}
\begin{split}
&P(K_1,\ldots,K_L)=\frac{1}{Z}\exp\left[\sum_t \left( h(K_t)
+\log \binom{N}{K_t}\right)\right.
\\& \quad \left. +\sum_{t}\sum_{u=1}^v J_u(K_t,K_{t+u})\right],
\end{split}
\eeq
where the binomial factors $\binom{N}{K_t}$ counts the spiking patterns $(\sigma_{1,t},\ldots,\sigma_{N,t})$ having $K_t$ spiking cells among $N$.

\section{Gaussian approximation}
It is possible to further simplify the maximum entropy model by treating $K$ as a continuous variable and constraint only its first and second moments $\<K_t\>$, $\<K_tK_{t+u}\>$. Using Eq.~\ref{eq:maxent}, these constraints lead to a Gaussian distribution for the number of spiking neurons:
\beq
P(K_1,\ldots,K_L)=\frac{1}{Z}\exp\left[-\frac{1}{2}\sum_t \sum_{u=0}^v x_t A_u x_{t+u}\right],
\eeq
where $x_t=K_t-\<K\>$ as before.
This process is equivalent to a generalized auto-regressive model:
\beq
x_t=\sum_{u=1}^v \gamma_u x_{t-u} + \epsilon_t,
\eeq
with $\epsilon_t$ a Gaussian variable of zero mean and covariance $\<\epsilon_t\epsilon_{t'}\>=\sigma^2\delta_{tt'}$ and
the correspondance:
\bea
A_0&=&\frac{1}{\sigma^2}\left(1+\sum_{u=1}^v \gamma_u^2\right)\\
A_u&=&-\frac{2}{\sigma^2}\left(\gamma_u-\sum_{|u'-u''|=u} \gamma_{u'}\gamma_{u''}\right).
\eea
This class of models generalizes Eq.~\ref{eq:autoreg1}. They predict a Gaussian distribution for the number of spiking neurons, in contradiction with experimental observations.

\section{Model solution}
The fitting problem of the maximum entropy distribution reduces to finding the parameters $h(K)$, $J_u(K,K')$ so that the distribution in Eq.~\ref{eq:Kmodel}
agrees with the experiments on the values of the marginal probabilities $P_u(K,K')$ (for all $K,K'$). Data estimates are simply obtained from the frequency of $(K_t,K_{t+u})$ pairs in the recordings. The model prediction, defined by Eq.~\ref{eq:pairwise}, requires to sum over all possible trajectories of $K_t$, which, if done with brute force, would be prohibitively long. However, it is possible to perform these sums using the technique of transfer matrices, which requires much less computational power. This technique is commonly used to solve one-dimensional problems in statistical mechanics. It is also known in computer science as an instance of dynamic programming.

\begin{figure}[t]
\begin{center}
\noindent\includegraphics[width=.79\linewidth]{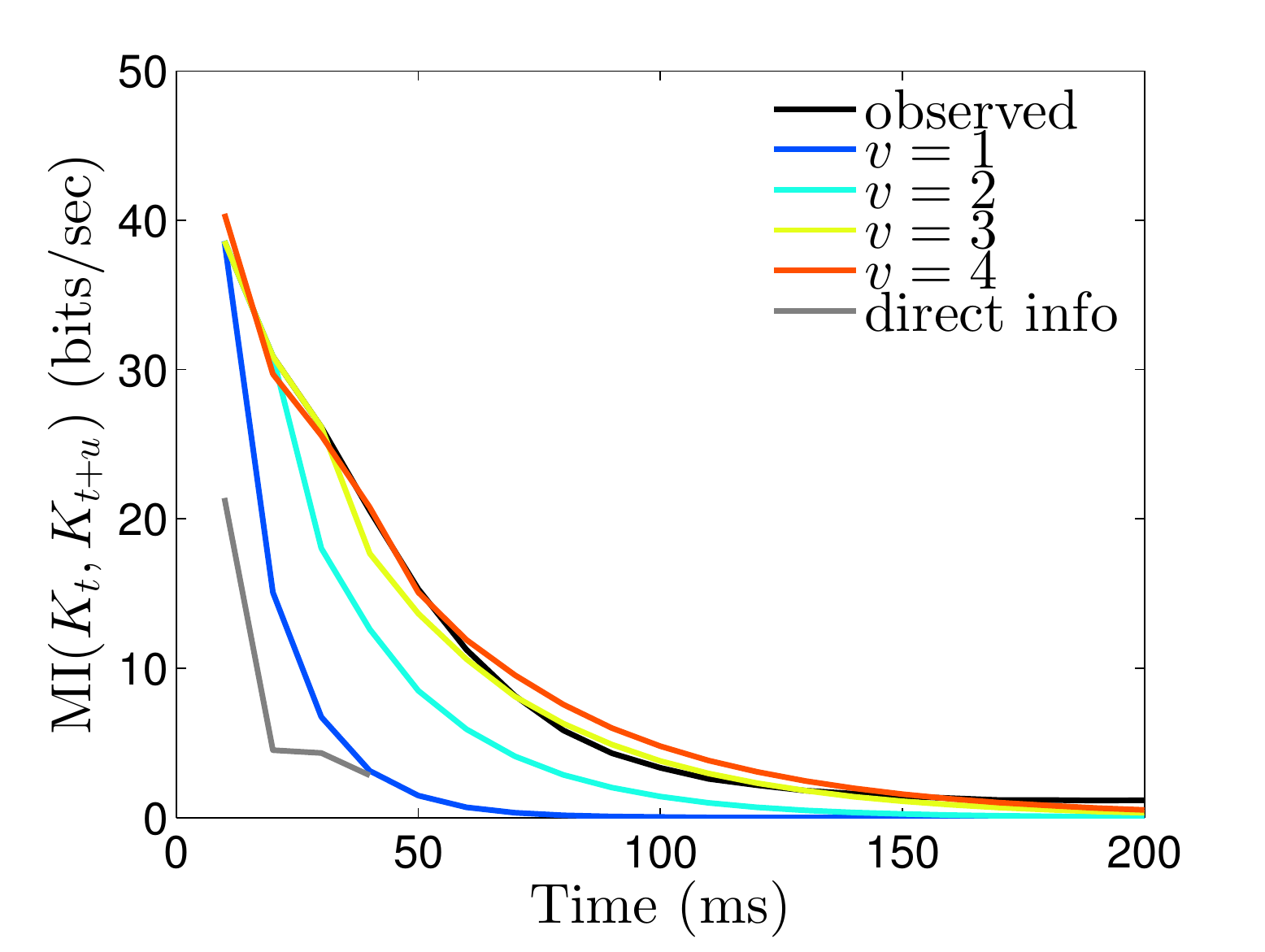}
\caption{{\bf Temporal correlations.} Mutual information between $K_t$ and $K_{t+u}$ as a function of $u\times \Delta t$ ($\Delta t=10$\,ms), for all $N=185$ neurons. The mutual information quantifies the correlation between two quantities. The model prediction for different $v$ is compared to the data. The agreement is good for $v=3$ and $4$. The gray curve shows the direct information between different times \cite{Weigt}, which quantifies the strengh of interaction between $t$ and $t+u$, within the $v=4$ model.
\label{fig:MI}
}
\end{center}
\end{figure}

\begin{figure}[t]
\begin{center}
\noindent\includegraphics[width=.99\linewidth]{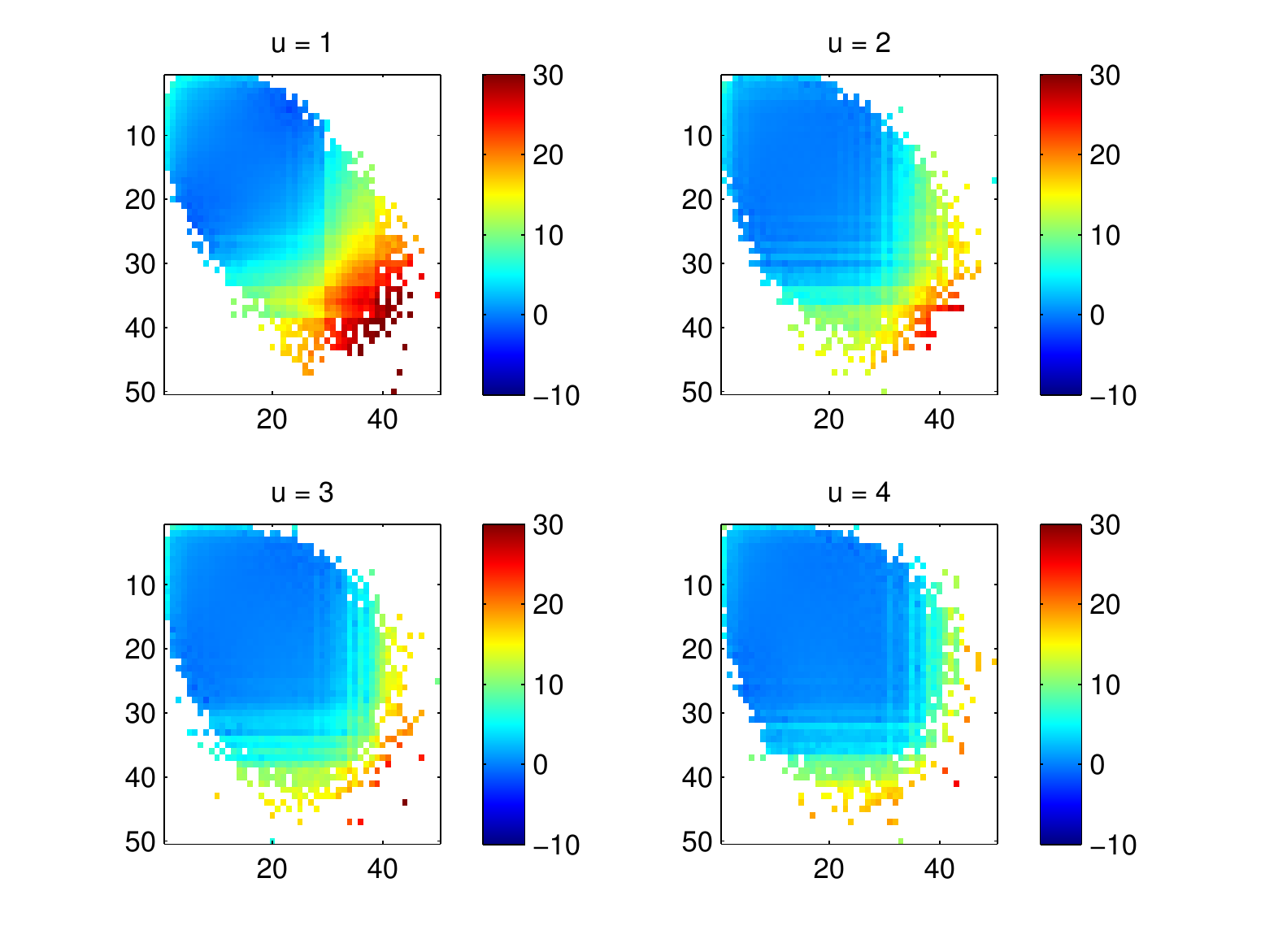}
\caption{{\bf Value of the coupling parameters $J_u(K_t,K_{t+u})$.} The x and y axes represent $K_t$ and $K_{t+u}$, respectively. The model was fittted with $v=4$ and all $N=185$ neurons.
\label{fig:J}
}
\end{center}
\end{figure}

We start by assuming that the trajectory $(K_1,\ldots,K_L)$ is an
$v^{\rm th}$ order Markov process (this assumption will be verified
later). We define the super variable
$X_t=(K_t,K_{t+1},\ldots,K_{t+v-1})$, and rewrite Eq.~\ref{eq:Kmodel}
as:
\beq\label{eq:Xmodel}
\begin{split}
P(\{X_t\})=&\frac{1}{Z}\exp\left[\sum_{t} H(X_t)+\sum_{t} W(X_t,X_{t+1})\right]
\\
&\times \prod_t \prod_{u=1}^{v-1} \delta_{X_t^{(u)},X_{t-1}^{(u+1)}}
\end{split}
\eeq
where $X_t^{(u)}$ is the $u^{\rm th}$ component of $X_t$, {\em i.e.} $K_{t+u-1}$, and
with
\beq
\begin{split}
H(X_t)=&\frac{1}{v}\sum_{u=1}^{v} \left[h(K_{t+u-1}) + \log\binom{N}{K_{t+u-1}}\right] \\
&+ \sum_{u'<u=1}^v \frac{1}{v-(u-u')} J_{u-u'}(K_{t+u-1},K_{t+u'-1}).
\end{split}
\eeq
and
\beq
W(X_t,X_{t+1})=J_v(K_t,K_{t+v}).
\eeq
If $K_t$ is $v^{\rm th}$-order Markovian, then the super-variable $X_t$ is Markovian:
\beq\label{eq:Markov}
P(\{X_t\})=P(X_1)\prod_{t=2}^L P(X_t|X_{t-1}).
\eeq
The conditional distribution can be written in the form:
\beq\label{eq:Xcond}
\begin{split}
P(X_t|X_{t-1})&=\frac{1}{z^t_\to}e^{H(X_t)+g_{\to}^t(X_t)-g_{\to}^{t-1}(X_{t-1})+W(X_{t-1},X_{t})}\\
&\times \prod_{u=1}^{v-1} \delta_{X_t^{(u)},X_{t-1}^{(u+1)}},
\end{split}
\eeq
where $g_{\to}^t$ is a function that will be specified by normalization (see below).
This identification can be verified by replacing Eq.~\ref{eq:Xcond} into Eq.~\ref{eq:Markov} and comparing with Eq.~\ref{eq:Xmodel}, with $g_\to^L(X_L)=0$,
\beq
P(X_1)=\frac{1}{z_1} e^{H(X_1)+g^1_\to(X_1)},
\eeq
 and 
$Z=\prod_{t=1}^L z_\to^t$. Thus, $X_t$ is indeed Markovian, and $K_t$
is $v^{\rm th}$ order Markovian.

The parameter to be learned is the function $g_{\to}(X)$. In general
that function depends on $t$, but here we assume that it is constant because of stationarity. This assumption is only valid in the bulk ({\em i.e.} for $t$ far away from both $1$ and $L$).
The normalization condition
\beq
\sum_{X_t} P(X_t|X_{t-1})=1,
\eeq
which must hold for all $X_{t-1}$, gives the following self-consistent equation for $g_{\to}(X)$:
\beq\label{eq:matrix}
e^{g_{\to}(X)}=\frac{1}{z}\sum_{X'} \left[e^{H(X')+W(X,X')}\prod_{u=1}^{v-1} \delta_{X'^{(u)},X^{(u+1)}}\right]e^{g_{\to}(X')},
\eeq
where we have replaced $X_{t-1}$ by $X$ and $X_{t}$ by $X'$ to ease notations (but also because these are dummy variables).
We can view this equation as an eigenvalue problem: $e^{g_{\to}(X)}$
is the eigenvector of the matrix defined in the bracket (called the
transfer matrix), associated with its largest eigenvalue $z$. This
equation can be solved by simply iterating Eq.~\ref{eq:matrix}, and
normalizing $e^{g_{\to}(X)}$ after each iteration (by {\em e.g.} $\max_X g_\to(X)$ or $\sum_X g_\to(X)$). After convergence, that normalization constant at each step simply gives the eigenvalue $z$. This procedure takes a computational time of order $(K_{\rm max}+1)^{v+1}$, which is large but manageable for small enough $v$.

The same reasoning can be repeated by writing the Markov dynamics of $X_t$ backward in time:
\beq\label{eq:Xcondback}
\begin{split}
P(X_t|X_{t+1})&=\frac{1}{z}e^{H(X_t)+g_{\gets}(X_t)-g_{\gets}(X_{t+1})+W(X_{t},X_{t+1})}\\
&\times \prod_{u=1}^{v-1} \delta_{X_t^{(u+1)},X_{t+1}^{(u)}},
\end{split}
\eeq
which gives the self-consistent equation:
\beq\label{eq:matrixback}
e^{g_{\gets}(X)}=\frac{1}{z}\sum_{X'} \left[e^{H(X')+W(X',X)}\prod_{u=1}^{v-1} \delta_{X'^{(u+1)},X^{(u)}}\right]e^{g_{\gets}(X')}.
\eeq
The only difference with Eq.~\ref{eq:matrix} is the exchange of $X$
and $X'$. Thus, $g_{\to}$ and $g_\gets$ may be different for general time-irreversible processes. The eigenvalue $z$ remains unchanged, however, because the right and left eigenvalues of a matrix are the same.

Armed with $g_{\to}$ and $g_{\gets}$, we can now calculate all marginals. Using
the Markovian nature of the sequence:
\beq
\begin{split}
P(\{X_{t'}\})=&P(X_t) P(X_1,\ldots,X_{t-1}|X_t)
P(X_{t+1},\ldots,X_L|X_t)\\
=& P(X_t) \prod_{t'=2}^t P(X_{t'-1}|X_{t'})
\prod_{t'=t}^{L-1} P(X_{t'+1}|X_{t'}),
\end{split}
\eeq
and replacing with Eqs.~\ref{eq:Xcond},\ref{eq:Xcondback} and
\ref{eq:Xmodel}, we get:
\beq
P(X_t)=\frac{1}{z_t} e^{g_{\to}(X_t)+g_{\gets}(X_t)+H(X_t)}
\eeq
and
\beq
\begin{split}
&P(X_t,X_{t+1})=P(K_t,\ldots,K_{t+v}) \\
&\quad =
\frac{1}{z_t z_{t+1}} e^{g_{\to}(X_{t+1})+g_{\gets}(X_t)+H(X_t)+H(X_{t+1})+W(X_{t},X_{t+1})}.
\end{split}
\eeq
We can also calculate pairwise marginals between $K_t$ at arbitrary
time differences by using the following recursion, for $u>v$:
\begin{widetext}
\beq\label{eq:recursion2points}
P(K_{t},K_{t+u+1},\ldots,K_{t+u+v})=
\sum_{K_{t+u}}
P(K_{t},K_{t+u},\ldots,K_{t+u+v-1}) P(K_{t+u+v}|K_{t+u},\ldots,K_{t+u+v-1}).
\eeq
\end{widetext}
starting with $u=0$:
\beq
P(K_{t},K_{t+1},\ldots,K_{t+v})=P(X_{t},X_{t+1}).
\eeq

This whole procedure can be performed at an arbitrary inverse temperature
$\beta$. 
The energy of a given spike train is, according to Eq.~\ref{eq:model}:
\beq\label{eq:energySI}
E=-\sum_t h(K_t)-\sum_{t}\sum_{u=1}^v J_u(K_t,K_{t+u}),
\eeq
and thus at temperature $1/\beta$ the distribution of spike trains reads:
\beq
P_\beta(\{\sigma_{i,t}\})=\frac{1}{Z(\beta)}e^{-\beta E(\{K_{t}\})},
\eeq
where $Z(\beta)$ enforces normalization.
The distribution $P_\beta(K_1,\ldots,K_t)$ is given by Eq.~\ref{eq:Kmodel} with the substitutions:
\bea
h(K)&\to& \beta h(K),\\
J_u(K,K') &\to & \beta J_u(K,K').
\eea
All the
results of the procedure, $z(\beta)$, $g_\to(X;\beta)$ and
$g_\gets(X;\beta)$ thus depend on $\beta$.
The free energy $F(\beta)=-\beta^{-1}\log Z(\beta)$ can be calculated per
unit time through $f(\beta)\equiv F(\beta)/NL=-\beta^{-1} \log z(\beta)/N$.
The average energy (Eq.~\ref{eq:energySI}) per unit time is given by:
\beq
\begin{split}
\epsilon(\beta)\equiv &\frac{\<E\>_\beta}{NL}=-\frac{1}{N}\sum_{K_t} h(K_t)P_\beta(K_t)\\
&-\frac{1}{N}\sum_{u=1}^v \sum_{K_t,K_{t+u}} J_u(K_t,K_{t+u})P_{\beta;u}(K_t,K_{t+u})
\end{split}
\eeq
and the entropy per unit time by $s(\beta)\equiv S(\beta)/NL=\beta \epsilon(\beta)+\log z(\beta)/N$. The specific heat $c(\beta)=-\beta \partial s/\partial \beta$ is obtained by numerical derivation.

The technique of transfer matrices can also be extended to calculate
the statistics of avalanches. Two distributions can be calculated:
that of the duration of the avalanche, and that of the number of
spikes in it. An avalanche starts at $t$ if $K_{t-1}=0$ and
$K_{t}>0$. It ends after $\ell$ steps if $K_{t+\ell}=0$, and
$K_{t'}>0$ for all $t'$ such that $t\leq t'<t+\ell$. The probability $Q_\ell$
for an avalanche to last at least $\ell$ steps, and have $K_{t+\ell},\ldots,K_{t+\ell+v-1}$
spiking neurons at the $v$ subsequent step is given recursively by:
\begin{widetext}
\beq
Q_\ell(K_{t+\ell},\ldots,K_{t+\ell+v-1})=\sum_{K_{t+\ell-1}>0}
Q_\ell(K_{t+\ell-1},\ldots,K_{t+\ell+v-2}) P(K_{t+\ell}|K_{t+\ell-1},\ldots,K_{t+\ell+v-2})
\eeq
with initialization $\ell=0$:
\beq
Q_\ell(K_{t},\ldots,K_{t+v-1})=\frac{P(K_{t-1}=0,K_{t},\ldots,K_{t+v-1})}{P(K_{t-1}=0)}=\frac{P(X_{t-1},X_t)}{P(K_{t-1}=0)}.
\eeq
Then the probability that the avalanche lasts $\ell$ steps is
calculated through:
\beq\label{eq:marg}
P_{\ell} = \sum_{K_{t+\ell+1},\ldots,K_{t+\ell+v-1}} Q_\ell(K_{t+\ell}=0,K_{t+\ell+1},\ldots,K_{t+\ell+v-1}).
\eeq
Restricting to non-zero avalanches, the distribution is given by ${P_{\ell}}/({1-P_{\ell=0}})$.

The distribution of the number of spiking events in the avalanche can
be calculated in a similar way, although at a higher computational
cost. We define $R_\ell(K_{t+\ell},\ldots,K_{t+\ell+v-1};n)$ as the
probability that an avalanche has lasted at least $\ell$ steps, 
has accumulated $n$ spiking events during these steps, and has
$(K_{t+\ell},\ldots,K_{t+\ell+v-1})$ spiking cells in the $v$
time windows following the $\ell^{\rm th}$ step. Then the
following recursion holds:
\beq
R_\ell(K_{t+\ell},\ldots,K_{t+\ell+v-1};n)=\sum_{K_{t+\ell-1}>0}
R_\ell(K_{t+\ell-1},\ldots,K_{t+\ell+v-2};n-K_{t+\ell-1}) P(K_{t+\ell}|K_{t+\ell-1},\ldots,K_{t+\ell+v-2}).
\eeq
The initialization at $\ell=0$ simply reads:
\beq
R_\ell(K_{t},\ldots,K_{t+v-1},n)=\frac{P(K_{t-1}=0,K_{t},\ldots,K_{t+v-1})}{P(K_{t-1}=0)} \delta_{n,0}
\eeq
As before the joint distribution $P_{\ell,n}$ for the size and
duration of avalanches is obtained by summing
over $K_{t+\ell+1},\ldots,K_{t+\ell+v-1}$ as in Eq.~\ref{eq:marg}, and
restricting to non-zero avalanches ($\ell>0$).

\end{widetext}

\begin{figure*}
\begin{center}
\noindent\includegraphics[width=.32\linewidth]{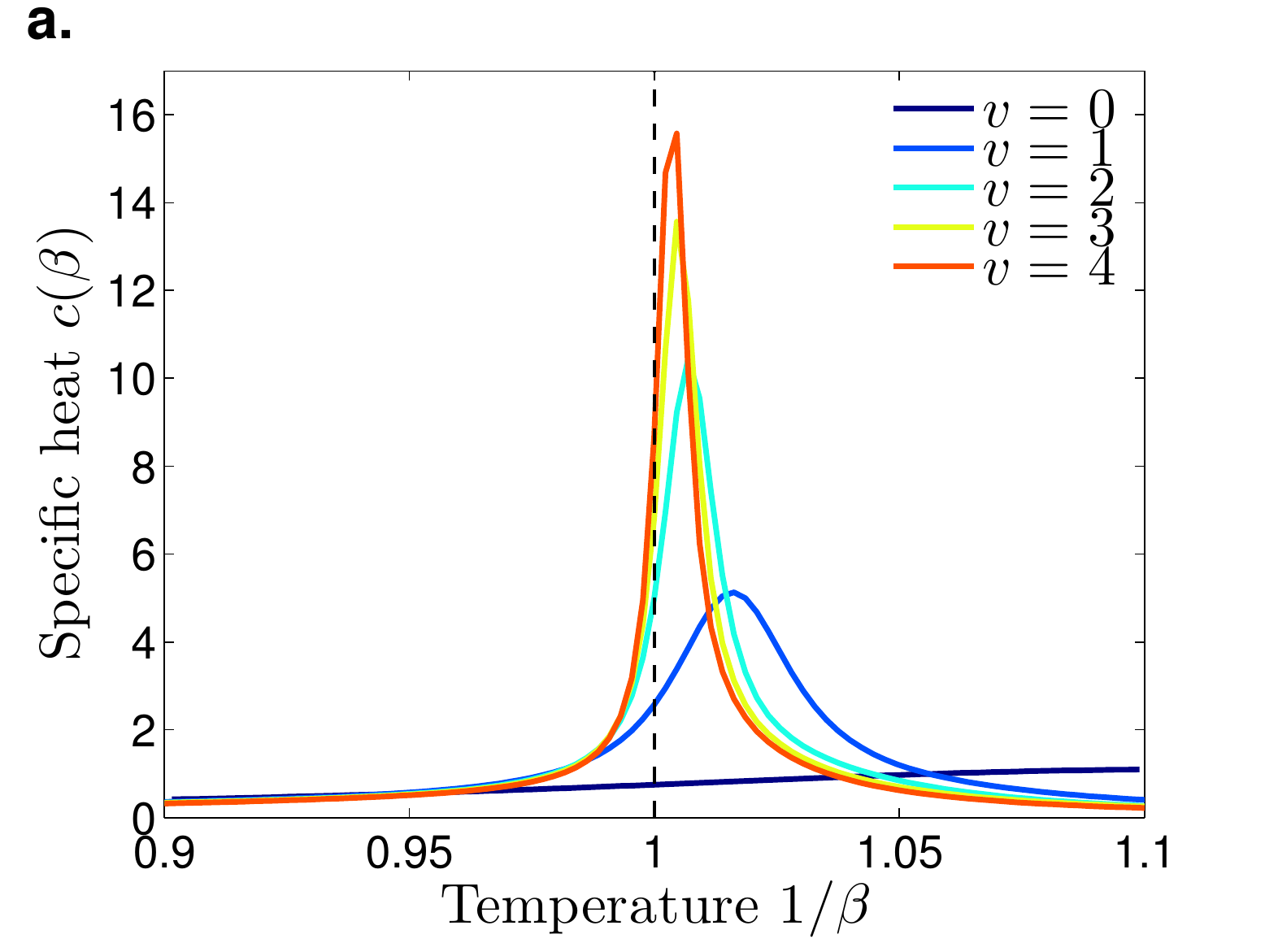}
\noindent\includegraphics[width=.32\linewidth]{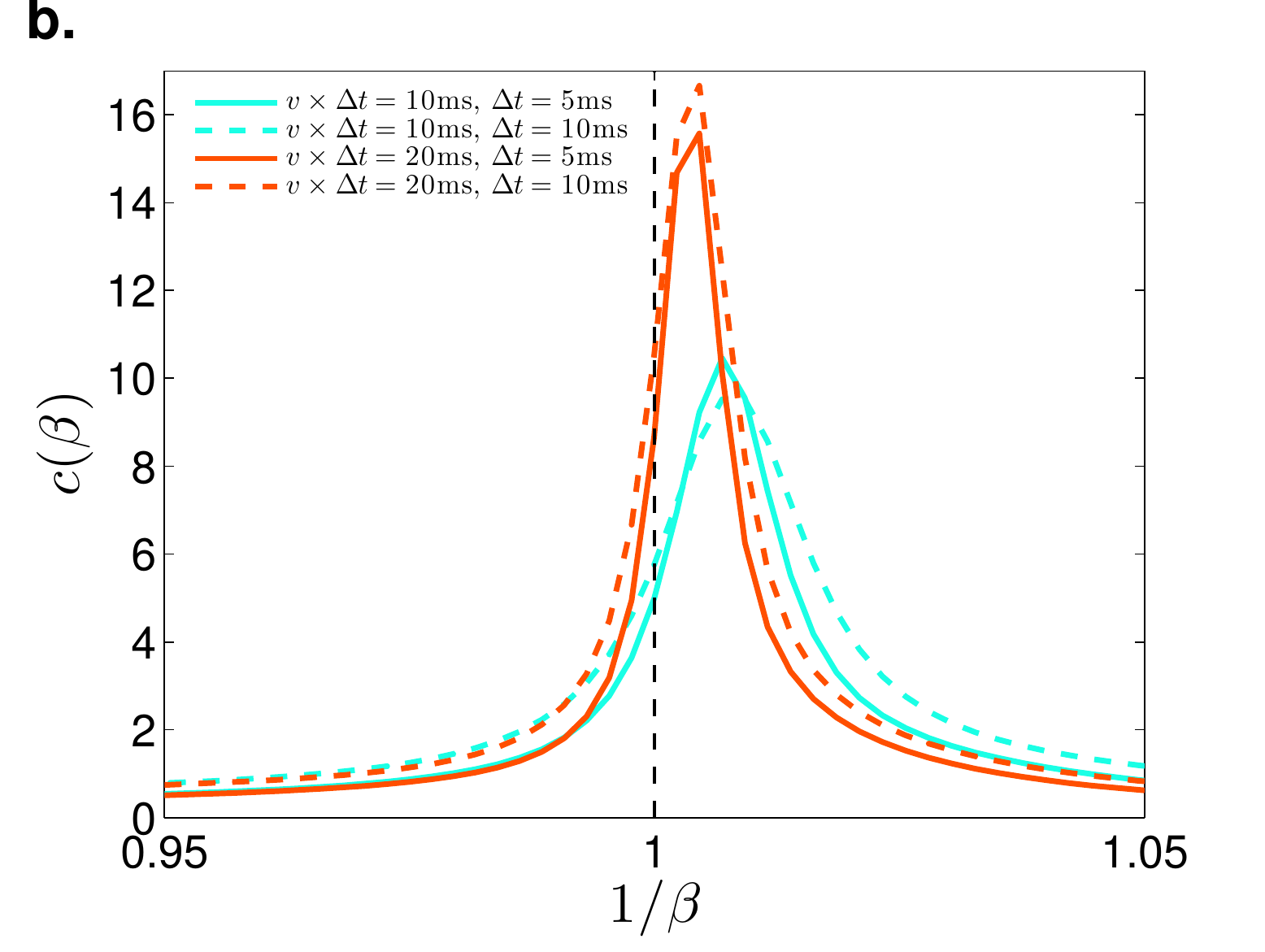}
\noindent\includegraphics[width=.32\linewidth]{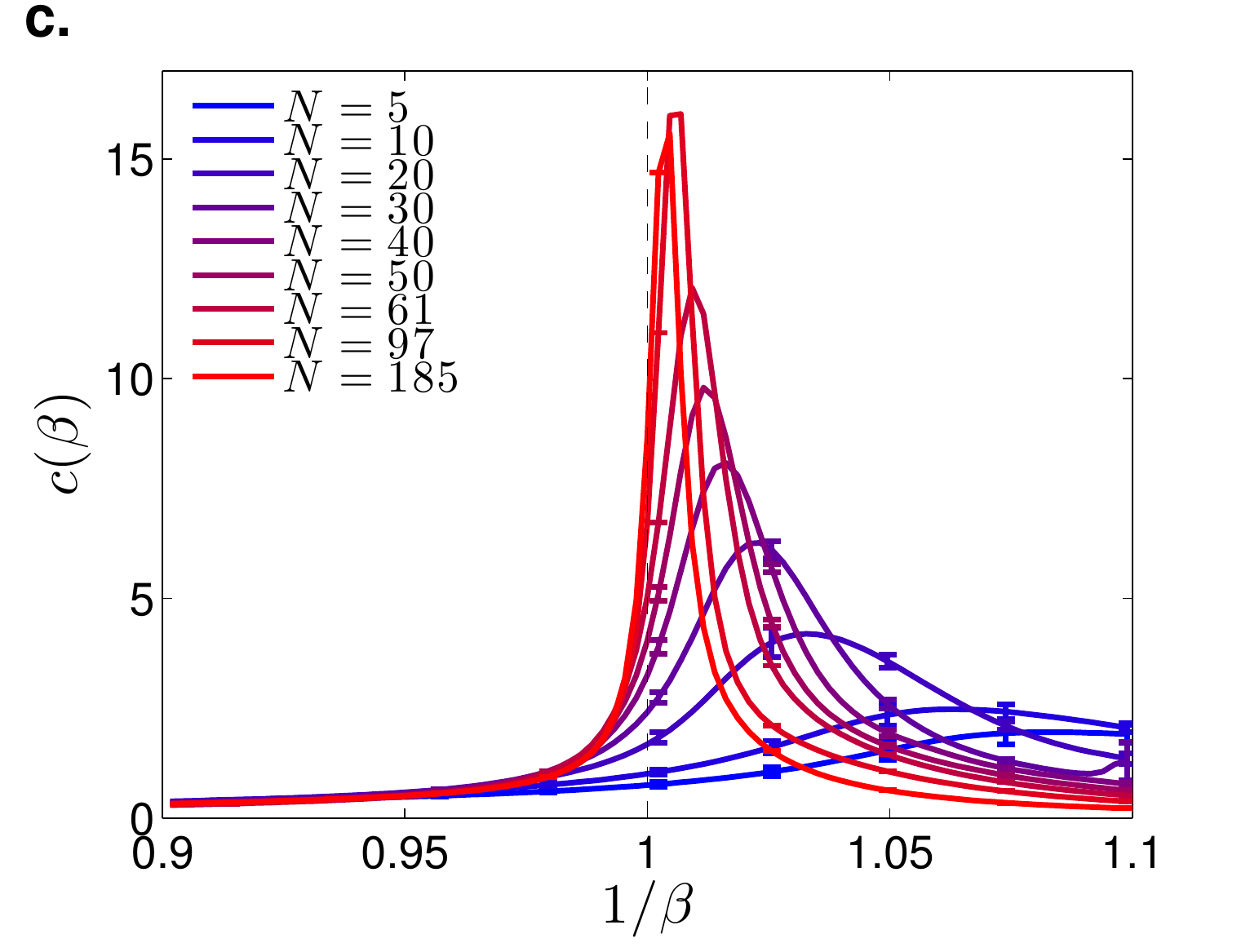}
\caption{{\bf Effect of window size on the specific heat.} Same as figure 2 of the main text, for a window size $\Delta t=5$ ms. (a)
Specific heat $c(\beta)$ of spike trains of the entire population
($N=185$), as a function of temperature $1/\beta$, for an increasing
temporal range $v$. (b) Comparison with the curves obtained for
$\Delta t=10$\,ms, with the same temporal range $v\times \Delta t$
expressed in seconds (=10\,ms for cyan curves, 20\,ms for the red
curves). Solid lines are for $\Delta t=5$\,ms, and dashed line for $\Delta t=10$\,ms. (c) Specific heat of spike trains of subnetworks of increasing sizes $N$, for $v=4$. Each point is averaged over 100 random subnetworks for $N\leq 50$, and shows one representative network for $N=61$ and $97$. The error bars show standard deviations.
\label{fig:specheat5}
}
\end{center}
\end{figure*}

\begin{figure}
\begin{center}
\noindent\includegraphics[width=.49\linewidth]{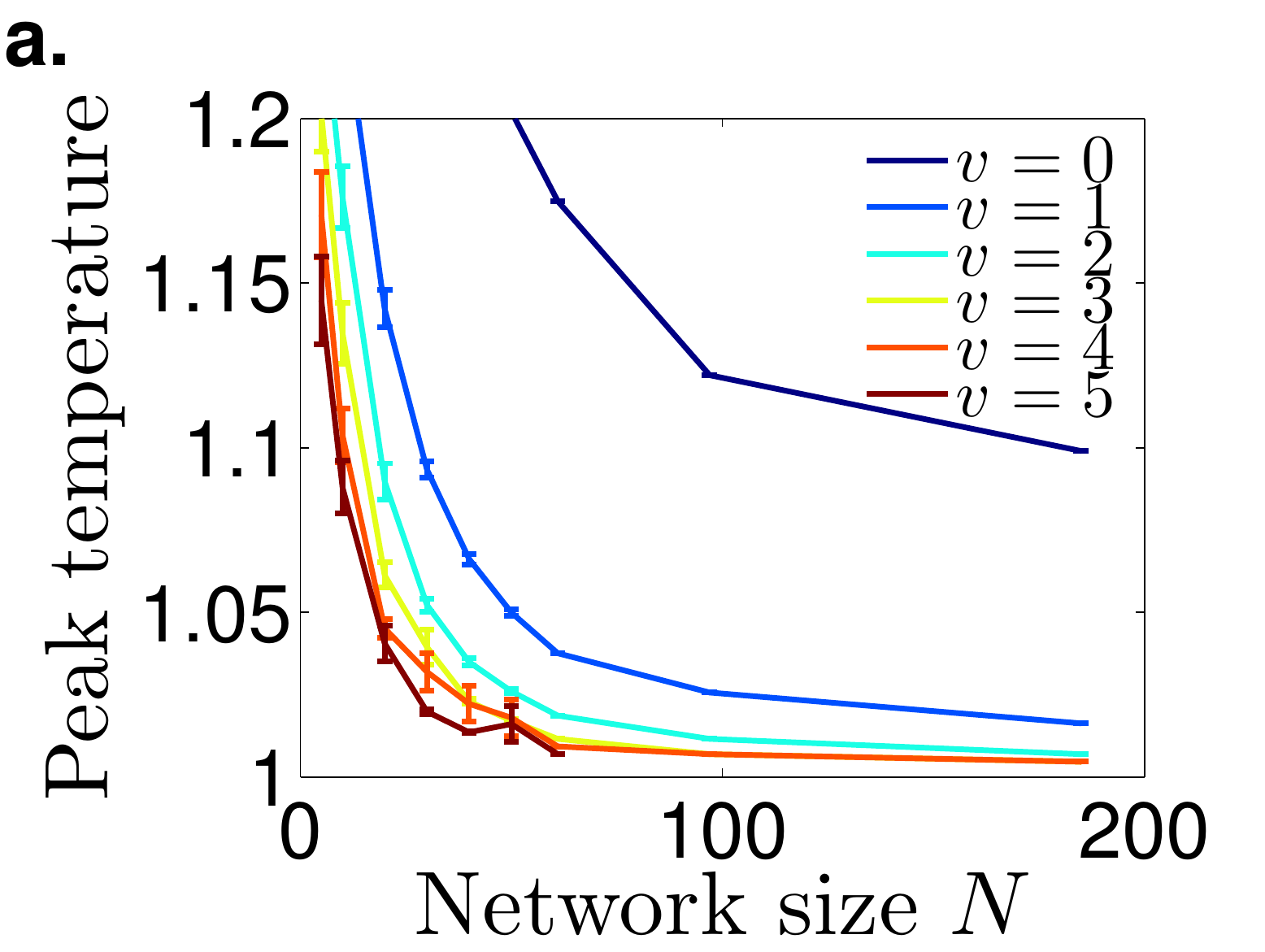}
\noindent\includegraphics[width=.49\linewidth]{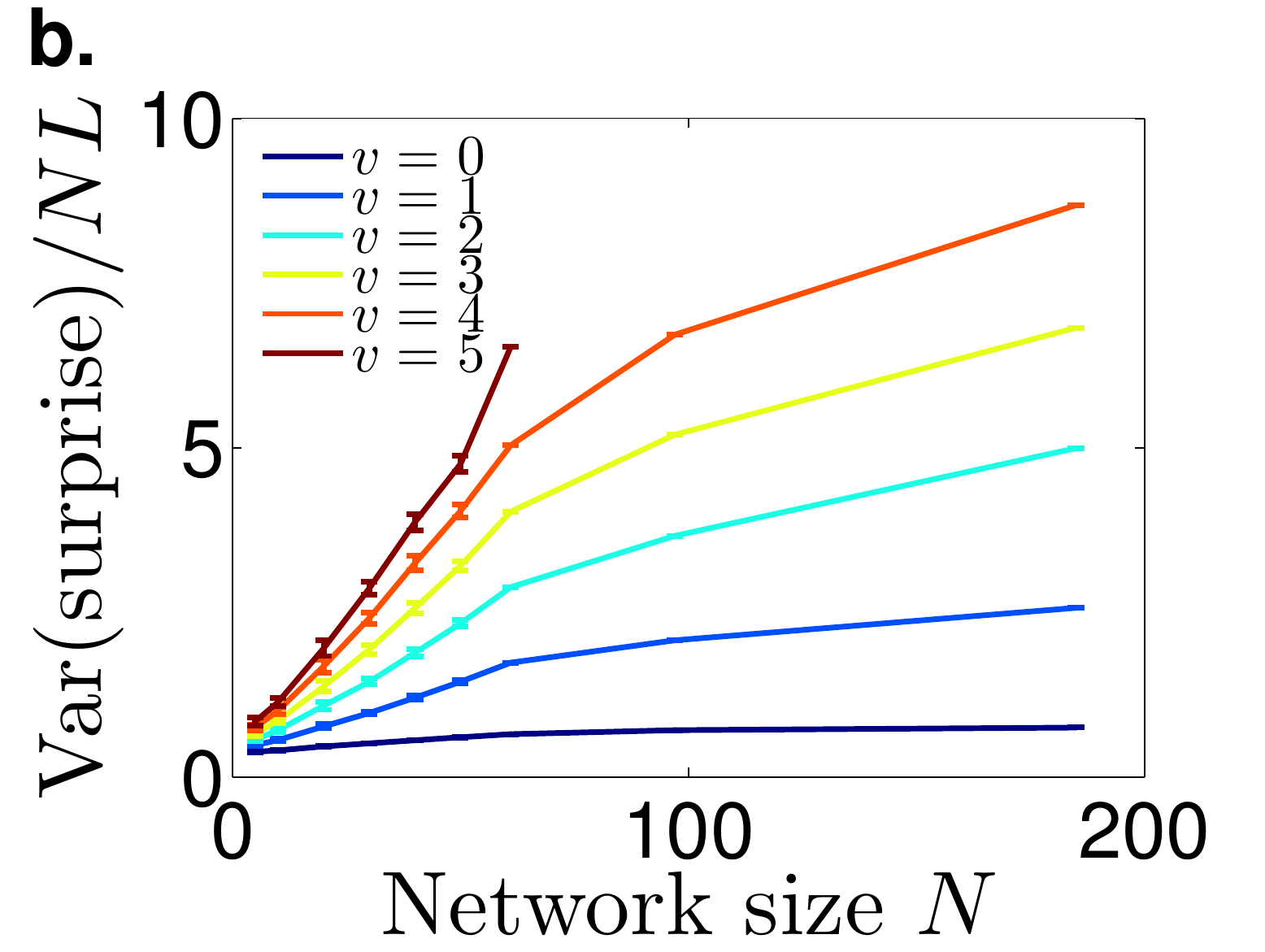}
\caption{{\bf Effect of window size on finite-size scaling.} Same as figure 3 of the main text, for a window size $\Delta=5$ ms. (a) Position of the peak  $1/\beta$ in specific heat  as a function of network size $N$, for an increasing temporal range $v$. (b) Normalized variance of the surprise as a function of $N$, for an increasing temporal range $v$.
\label{fig:surprise5}
}
\end{center}
\end{figure}

\section{Model learning}
The procedure described in the previous section allows us to calculate the marginals and thermodynamic quantities for a given set of parameters $h(K)$ and $J_u(K,K)$. We want to solve the inverse problem, which is to find these parameters for a given set of marginals $P_u(K_t,K_{t+u})$. To do this we implement the following iteration:
\begin{widetext}
\bea
h(K) &\gets & h(K)+\epsilon\left[ P_{\rm data}(K)-P_{\rm model}(K)\right]\label{eq:step1}\\
J_u(K_t,K_{t+u}) &\gets & J_u(K_t,K_{t+u})+\epsilon\left[ P_{\rm data}(K_t,K_{t+u})-P_{\rm model}(K_t,K_{t+u})\right],\label{eq:step2}
\eea
after which we enforce our constraints (Eqs.~\ref{eq:C1}, \ref{eq:C2}, \ref{eq:C3}) by:
\bea
h(K) & \gets & h(K) + \sum_{u=1}^v\left[\sum_{K'} P(K') J_u(K,K') + \sum_{K'} P(K') J_u(K',K)\right]\\
h(K) &\gets & h(K)-\sum_{K'} P(K')h(K')\\
J_u(K,K')&\gets & J_u(K,K')-\sum_{K''} P(K'') J_u(K'',K')-\sum_{K''} P(K'') J_u(K,K'')+ \sum_{K'',K'''} P(K'')P(K''') J_u(K'',K''')
\eea
\end{widetext}
Note that only the first step \ref{eq:step1},\ref{eq:step2} actually modifies the model. At each step, $P_{\rm model}$ must be re-calculated from the new set of parameters $(h,J_u)$.

We initialize the algorithm by setting $J_u=0$ for $u>1$. This corresponds to the case $v=1$, for which $h(K)$ and $J_1(K,K')$ can be deduced directly from $P(X_t|X_{t-1})$.

This procedure is equivalent to a gradient descent algorithm on the log-likelihood \cite{Mora:2011p12729}, and therefore is guaranteed to converge to the solution provided that $\epsilon$ is small enough.

\section{Inferred parameters, and choice of $v$}
To assess the performance of the model, we can ask how well it predicts the correlations of $K$ at different times. The mutual information, defined as:
\beq
\mathrm{MI}(K_t,K_{t+u})=\sum_{K,K'} P_u(K,K')\log \frac{P_u(K,K')}{P(K)P(K')},
\eeq
is a non-parametric measure of these correlations. Fig.~\ref{fig:MI} shows this mutual information estimated from the data, as well as its prediction for models with different $v$. Note that by construction, the agreement is perfect for $u\leq v$. The $v=3$ and $v=4$ model predictions are fairly good even for larger $u$, indicating that a larger $v$ would not improve the model prediction much.

The inferred $J_u(K_t,K_{t+u})$ are represented in Fig.~\ref{fig:J} for $v=4$ and $N=185$. They become smaller as $u$ increases, indicating that the effective interactions between time windows decay with the time difference. This can be quantified using the Direct Information, which measures the strength of interaction between two variables in a complex interaction network \cite{Weigt}. The direct pairwise distribution is defined as:
\beq
P^{\rm dir}_u(K,K')=e^{J_u(K,K')+\phi(K)+\phi'(K')},
\eeq
where $\phi(K)$ and $\phi(K')$ are chosen so that $\sum_K P^{\rm dir}_u(K,K')=P(K')$, and $\sum_{K'} P^{\rm dir}_u(K,K')=P(K)$. This distribution corresponds to the effect that $K_t$ and $K_{t+u}$ would have on each other if they were not interacting with $K_{t'}$ at other times $t'$.
The direct information is then defined as the mutual information in this pairwise distribution:
\beq
\mathrm{DI}(K_t,K_{t+u})=\sum_{K,K'} P^{\rm dir}_u(K,K')\log \frac{P^{\rm dir}_u(K,K')}{P(K)P(K')}.
\eeq

This quantity is represented in gray in Fig.~\ref{fig:MI}, and shows a sharp decay as a function of $u$, a further indication that $v=4$ is sufficient.

\section{Effect of the window size}
Both the thermodynamic approach and the model used to describe spike trains depend on the window size $\Delta t$. We repeated the analysis with a shorter window size of $\Delta t=5$\,ms. The results are shown in Figs.~\ref{fig:specheat5} and \ref{fig:surprise5}.

In the limit of small window sizes, we expect that models with different $\Delta t$, but with the same temporal range in seconds,  $v\times \Delta t$, should yield similar predictions. Fig.~\ref{fig:specheat5}b shows that this is indeed the case. This indicates that the results of our analysis do not depend much on the choice of window size.

\end{document}